%% file: Manuscript.tex
\documentclass[aoas]{imsart}

\RequirePackage{amsthm,amsmath,amsfonts,amssymb}
\RequirePackage[authoryear]{natbib}
\RequirePackage[colorlinks,citecolor=blue,urlcolor=blue]{hyperref}
\RequirePackage{graphicx}
\usepackage{threeparttable} 
\RequirePackage{makecell}
\usepackage{hyperref}
\usepackage{float}
\usepackage{soul}
\usepackage{tabularray}
\usepackage{xcolor}
\usepackage{placeins}
\usepackage{lscape}
\usepackage{booktabs}
\usepackage{fix-cm}
\startlocaldefs

\numberwithin{equation}{section}

\theoremstyle{plain}

\newtheorem{theorem}{Theorem}[section]

\newtheorem{corollary}{Corollary}
\newtheorem{assumption}{Assumption}

\endlocaldefs

\begin{document}

\begin{frontmatter}

\title{Causal indirect effect of an HIV curative treatment: mediators subject to an assay limit and measurement error }

\runtitle{Causal indirect effect of an HIV curative treatment}

\begin{aug}

\author[A]{\fnms{Vindyani}~\snm{Herath}\ead[label=e1]{vindyani@bu.edu}},
\author[B]{\fnms{Ronald J}~\snm{Bosch}\ead[label=e2]{rbosch@hsph.harvard.edu}}
\and
\author[A]{\fnms{Judith}~\snm{Lok}\ead[label=e3]{jjlok@bu.edu}}

\address[A]{Department of Mathematics and Statistics,
Boston University\printead[presep={,\ }]{e1}}

\address[B]{Center for Biostatistics in AIDS Research, Harvard
T.H. Chan School of Public Health, Boston, MA\printead[presep={,\ }]{e2,e3}}
\end{aug}

\begin{abstract}
Causal mediation analysis decomposes the total effect of a treatment on an outcome into the indirect effect, operating through the mediator, and the direct effect, operating through other pathways. One can estimate only the pure indirect effect/indirect effect relative to no treatment, rather than the total effect by combining a hypothesized treatment effect on the mediator with outcome data without treatment. Furthermore, the mediation formula holds for the pure indirect effect (or the organic indirect effect relative to no treatment) regardless of whether there is an interaction between the treatment and mediator in the outcome model. This methodology holds significant promise in selecting prospective treatments based on their indirect effect for further evaluation in randomized clinical trials.

We apply this methodology to assess which of two measures of HIV persistence is a more promising target for future HIV curative treatments. We combine a hypothesized treatment effect on two mediators, and outcome data without treatment, to compare the indirect effect of treatments targeting these mediators. Some HIV persistence measurements fall below the assay limit, leading to left-censored mediators. We address this by assuming the outcome model extends to mediators below the assay limit and use maximum likelihood estimation. To address measurement error in the mediators, we adjust our estimates. 
Using data from completed ACTG studies, we estimate the pure or organic indirect effect of potential curative HIV treatments on viral suppression through weeks 4 and 8 after HIV medication interruption, mediated by HIV persistence measures.  

\end{abstract}


\end{frontmatter}

\section{Introduction}
Causal mediation analysis (\cite{Baron1986}, \cite{Imai2010}, \cite{Pearl_2001}, \cite{Vanderweele2015}) investigates the mechanisms that underlie an observed effect of an exposure variable on an outcome variable by examining their connection to a third intermediate variable: the mediator. Mediation analysis decomposes the total effect of a treatment on an outcome into the indirect effect, the effect that operates through a mediator, and the direct effect, the effect that operates through other pathways. Mediation analysis can serve as a methodological approach to assess the clinical and statistical significance of a hypothesized mechanism (\cite{Hardnett2009}). 

Several approaches to causal mediation analysis, defining direct and indirect effects, have been proposed (\cite{Robins1992}, \cite{Pearl_2001}). \cite{Lok2021} show that we can additionally use mediation analysis to estimate pure or organic indirect effects by combining a hypothesized effect on the mediator with outcome data obtained without intervention. This is valuable for identifying promising new treatments with favorable pure or organic indirect effects for further evaluation in randomized clinical trials. Organic indirect effects can be interpreted as the effect of the mediator being affected by the treatment and then the outcome following its course as though the resulting mediator value came about under no treatment (\cite{Lok2021}). Both pure and organic indirect effects are typically identified through the mediation formula (\cite{Pearl_2001}), leading to numerically identical results.

We use causal mediation analysis to evaluate which of two measures of HIV viral persistence is a more promising target for future HIV curative treatments, using a hypothesized effect of the HIV curative treatment on the mediator. 

HIV, the Human Immunodeficiency Virus, affects approximately 40 million people worldwide (\cite{unaids2024Global}). Even though antiretroviral therapy (ART) successfully suppresses HIV replication in most HIV-infected individuals, it is not a cure. Furthermore, individuals with HIV infection must take ART indefinitely and are still at increased risk for complications (\cite{Deeks}, \cite{matza2017risks}). To evaluate the clinical efficacy of new curative treatments, HIV curative treatment trials require an analytic treatment interruption (ATI). ATI studies temporarily discontinue ART to measure the time until the HIV viral load reaches a pre-defined threshold, the time of viral rebound. However, ATI studies are expensive due to frequent monitoring, and interrupting ART carries risks to study participants as it usually results in viral rebound within a few weeks. Hence, it is beneficial to select the most promising therapies for ATI trials, and they must be carried out with utmost caution. \cite{Li2016} used data from ATI studies and discovered that biomarkers indicating a larger HIV viral reservoir while on ART are associated with a shorter time to viral rebound. 

\cite{Lok2021} estimated the pure or organic indirect effect of a hypothetical new HIV curative treatment that leads to a ten-fold reduction of HIV persistence measures. They estimated the pure or organic indirect effect of a curative HIV treatment on viral suppression through ATI week 4 and week 8, mediated by two HIV persistence measures. They used data from participants in AIDS Clinical Trials Group (ACTG) studies who interrupted ART in the absence of curative HIV treatments.  Some HIV persistence measurements fall below the assay limit, resulting in a mediator that is left-censored.  \cite{Lok2021} assumed that how far below the assay limit a mediator lies does not affect the time to viral rebound. \cite{Chernofsky} extended these methods to settings where viral rebound depends on how far below the assay limit a mediator value lies. Consequently, \cite{Chernofsky} could incorporate larger treatment-induced reductions of the HIV viral reservoir.

However, none of these analyses accounted for measurement error, which is substantial for the HIV persistence measures considered. Measurement error arises from a combination of assay variation and short-term variation within individuals. While the underlying HIV persistence measures are relatively stable, daily and hourly variations and lab measurement errors contribute to deviations from the true level in single blood samples (\cite{Scully2022}). It is non-trivial to account for measurement error in mediators with an assay limit. Furthermore, this article shows that correcting for measurement error substantially impacts the results. 

This article estimates only the pure or organic indirect effect of a curative HIV treatment that reduces the HIV reservoir ten-fold or one hundred-fold (equivalently, shifts the distribution of the viral persistence measure on the log$_{10}$ scale by one-log$_{10}$ or two-log$_{10}$ to the left, respectively). That is, we hypothesize how a curative HIV treatment affects the distribution of the HIV persistence measure and then consider how ATI viral rebound would follow as though the viral persistence measure values resulted without curative intervention. This pure or organic indirect effect will often be the effect of the HIV curative treatment through its intended pathway, as HIV curative treatments often target viral persistence. We estimate only the pure indirect effect/indirect effect relative to no treatment, rather than the total effect. This choice aligns with the hypothesized pathway of interest. Furthermore, the mediation formula for the pure indirect effect (or organic indirect effect relative to no treatment) remains valid regardless of whether there is an interaction between the treatment and the mediator in the outcome model. Unlike the prior work by \cite{Chernofsky}, this study employs maximum likelihood estimation to assess the indirect effects, providing improved precision. It also addresses measurement error, which ensures more accurate and reliable estimates. The Delta method is used to calculate confidence intervals. Furthermore, extending \cite{Chernofsky}, this article presents results to evaluate the reduction in the viral reservoir required to substantially reduce the risk of viral rebound through both week 4 and week 8 after ATI. In addition, this analysis is conducted on a different population, improving the applicability of the results in the current era of HIV cure research, where ATI studies are restricted to individuals not on NNRTIs (non-nucleoside reverse transcriptase inhibitors) (\cite{Gunst2025}).

The remainder of this article is structured as follows. Section~\ref{section2} describes the HIV treatment interruption data from the AIDS Clinical Trials Group studies, which motivate the proposed methodology. Section~\ref{Methods} focuses on estimating pure or organic indirect effects when mediators are subject to an assay limit and measurement error. Sections~\ref{Results} and~\ref{Sim_section} provide the analysis results for the HIV treatment interruption data and the simulations evaluating the performance of our methods for small and large sample sizes and various shifts. Section~\ref{Discussion} discusses the findings. The Supplementary Appendix contains proofs and additional results. 

\section{Data} \label{section2}
This article analyzes data from six completed AIDS Clinical Trials Group (ACTG) studies: ACTG371, A5024, A5068, A5170, A5197, A5345, restricted to 104 participants living with HIV, not on NNRTI-based regimens (\cite{Li2016}, \cite{Li2022}). Participants on NNRTIs are excluded because of the long half-life of NNRTIs and their low barrier to resistance development (\cite{10.1093/cid/ciy463}). None of the 104 participants in these analytic treatment interruption (ATI) studies received immunologic interventions. 

Following \cite{Li2022}, viral rebound is defined as the time from ART interruption to the time that the first viral load in a participant’s blood is confirmed above 1000 copies/ml. The binary outcome variable is calculated as viral suppression through ATI week 4 (Yes if viral rebound was not observed by week 4; No otherwise). The binary outcome viral suppression through week 8  is also considered.

Two continuous viral persistence measures are considered as mediators: single-copy plasma HIV RNA (SCA) and cell-associated HIV RNA (caRNA). caRNA provides insights into the active viral population while SCA allows for the highly sensitive detection and quantification of HIV-1 RNA in plasma, even at very low levels. In the ACTG A5345 study, caRNA was measured with an assay limit of 28 copies per million CD4+ T cells, and SCA was measured with an assay limit of 0.56 copies per ml (\cite{Li2022}). In the remaining studies, caRNA was measured with an assay limit of 92 copies per million CD4+ T cells, and SCA was measured with an assay limit of 1 copy per ml (\cite{Li2016}). Both SCA and caRNA have measurement error, and interest lies in the effect mediated by their underlying values. The measurement error considered is a combination of short-term variation within individuals and assay variation. The standard deviation of the error for caRNA and SCA was estimated as 0.29 and 0.47 log$_{10}$, respectively, using data from completed ACTG studies (\cite{Scully2022}, \cite{Riddler2016}). 

Given that the mediator is not subject to randomization, for causal mediation analysis, it is essential to account for pre-treatment common causes: covariates that may affect both the mediator and the outcome (\cite{Vanderweele2015}). One factor that may influence both the HIV persistence measures and the timing of HIV viral rebound is the timing of ART initiation: studies have shown that individuals who initiated ART during acute or early infection tend to have lower viral reservoirs than individuals who initiated ART during chronic infection (\cite{Lietal_2024}). Therefore, whether participants initiated ART during acute/early HIV infection versus chronic HIV infection is considered as a common cause.

\section{Methodology} \label{Methods}
\subsection{Pure indirect effect / Organic direct and indirect effects relative to \texorpdfstring{$a = 0$}{a = 0} }

Let A be a binary treatment, Y the outcome of interest, M a mediator measured between A and Y, and C the pre-treatment common causes. 

The widely recognized natural indirect and direct effects have been defined using so-called cross-world counterfactual quantities  $Y^{(1, M^{(0)})}$ (\cite{Robins1992}, \cite{Pearl_2001}): the outcome under treatment had the mediator been set to the mediator value without treatment. The pure indirect effect
\begin{equation*}
    E\left[Y^{(0,M^{(1)})} \right] - E\left[Y^{(0)}\right]
\end{equation*}
is the difference between the expected outcome under treatment $a=0$, while varying the mediator from $M^{(0)}$, its value under $a=0$, to $M^{(1)}$, its value under $a=1$ (\cite{Robins1992}, \cite{Vanderweele2013}).

 \begin{figure}[ht]
\includegraphics[scale=0.6]{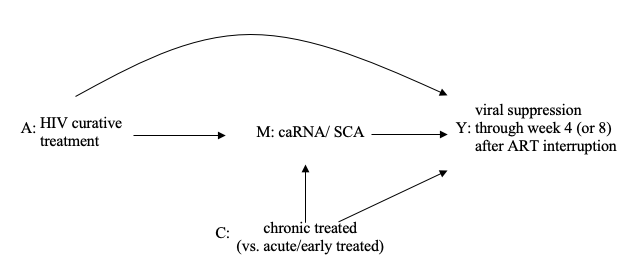}
\caption{Causal diagram for the effect of an HIV curative treatment that shifts the distribution of on-ART caRNA/SCA .}
\label{DAG}
\end{figure}

\cite{Lok2016} introduced the concept of ‘organic’ indirect and direct effects, which can be defined and estimated without the necessity of setting mediator values to specific values. Let $I$ be an intervention on the mediator. Denote $M^{(0,I=1)}$ and $Y^{(0,I=1)}$ as the mediator and outcome respectively, under $a=0$ and combined with intervention $I$ on the mediator. Consider organic indirect and direct effects relative to $a=0$. $I$ is defined to be an organic intervention relative to $a=0$ and C (\cite{Lok2021}) if
\begin{equation} \label{organic_IE_1}
    M^{(0,I=1)} \mid C = c \; \; \; \sim \; \; \; M^{(1)}  \mid C=c
\end{equation} and 
\begin{equation} \label{organi_IE_2}
    Y^{(0, I=1)} \mid M^{(0, I=1)} =m, C=c \; \; \; \sim \; \; \; Y^{(0)} \mid M^{(0)} = m, C=c.
\end{equation}

Combining organic interventions $I$ with “no treatment” allows one to consider the effect of an intervention $I$ that influences the mediator value M similarly to $a=1$ (\ref{organic_IE_1}), after which the outcome follows its natural course as though the mediator came about under no treatment $(a=0)$ (\ref{organi_IE_2}). \cite{Lok2016} showed that under randomized treatment, for an intervention $I$ that is organic relative to $a=0$ and $C$, the Mediation Formula ( \cite{Pearl2010},\cite{Pearl2011}) holds:
\begin{equation}
    \mathbb{E}\left[ Y^{(0, I=1)} \right] = \int_{(m,c)} \mathbb{E} [Y \mid M=m, C=c, A=0 ] f_{M \mid C=c, A=1} (m) f_C(c)dm \; dc. \label{med.formula}
\end{equation}
The organic indirect effect relative to $a=0$ and $C$ is defined as  
\begin{equation*}
     \mathbb{E}[Y^{(0, I=1)}] - \mathbb{E}[Y^{(0)}].
\end{equation*}

Under the usual identifiability conditions, the pure indirect effect is identified by the same Mediation Formula (\cite{Vanderweele2015}). The pure indirect effect and organic indirect effect relative to $a=0$ depend on $a=1$ only through the distribution of the mediator. If the impact of the treatment on the distribution of the mediator is known or hypothesized, the indirect effect can be expressed as a function of the distribution of observations under $a=0$ alone. 

As a consequence, the pure or organic indirect effect $\mathbb{E}[Y^{(0, I=1) } ] - \mathbb{E}[Y^{(0) }]$ can be used to explore the reduction in the viral reservoir needed to significantly increase the probability of ATI viral suppression through weeks 4 and 8 in the absence of the direct effect. 
This article estimates the pure or organic indirect effect of an HIV curative treatment hypothesized to reduce the two viral persistence measures, given pre-treatment common causes, using existing ART interruption data under no curative treatment. 

\subsection{Statistical Analysis} \label{stat_analysis}
To estimate the pure or organic indirect effect of potential curative HIV treatments, we model the probability of viral suppression through ATI weeks 4 and 8, given the pre-treatment common causes and viral persistence measures on the log$_{10}$ scale. For participants $i=1,\ldots,N$,  the outcome viral rebound is categorized as suppressed $(Y_i=1)$ versus not suppressed $(Y_i=0)$ through weeks 4 and 8 of ART interruption. 

The primary analysis assumes a probit regression model for the outcome viral suppression. The probit model differs from the logistic regression model in that it uses the probit link instead of the logit link and is widely used to model binary outcomes in fields such as econometrics, public health, and the social sciences. A sensitivity analysis assumes a logistic regression model. 

\begin{assumption} \label{assump1}
    Assume the following model for the response variable $Y$ under no curative treatment:
    \begin{equation}\label{eq2}
        P(Y=1 \mid A = 0 , M=m, C=c) = \Phi(\beta_0 + \beta_1m + \beta_2c),
    \end{equation}
    where $\Phi(.)$ is the cumulative distribution function of the standard normal distribution. 
\end{assumption} 

\noindent Equation~(\ref{eq2}) can be written in the form 
\begin{equation*}
    P(Y=1\mid A=0,M=m,C=c)= P(Z \leq \beta_0 + \beta_1 m +\beta_2 c),
\end{equation*} where $Z$ is a standard normal random variable. 
In the HIV application, the true mediators $M_i$ are measured with error, resulting in observed mediator measurement values $M_i^\ast$. 

\begin{assumption}\label{assump2}
    For $i = 1,\ldots, N,$ let $M_i$  be the true mediator and assume that given $C_i=c, M_i$ follows a normal distribution with mean $\alpha_0 + \alpha_1 c$ and variance $\sigma_M^2$ that does not depend on $c$. That is, 
    \begin{equation*}
        M_i \mid C_i = c \; \sim \; \mathcal{N} (\alpha_0 + \alpha_1c, \sigma_M^2).
    \end{equation*}
\end{assumption}

\begin{assumption} \label{assump3}
    Assume the following measurement error model:
    \begin{equation*}
        M_i^\ast = M_i + U_i,
    \end{equation*}
    where the error $U_i$ is independent of $(Y_i, M_i, C_i)$, has mean $0$ and variance $\sigma_U^2$, and is normally distributed: $ U_i \sim \mathcal{N}(0 , \sigma_U^2)$. 
\end{assumption}

\noindent Assumptions \ref{assump2} and \ref{assump3} imply that given $C_i=c$, $M_i^\ast$ follows a normal distribution: 
\begin{equation*}
    M_i^\ast \mid C_i = c \sim \mathcal{N}(\alpha_0 + \alpha_1c , \sigma_{M^\ast}^2), 
\end{equation*}
with $\sigma_{M^\ast}^2 = Var(M_i^\ast \mid C_i ) = \sigma_M^2 + \sigma_U^2.$

To account for measurement error in the mediators, SCA and caRNA, estimates are adjusted following \cite{Valeri2014_ME}, using the probit regression model for the probability of viral suppression through ATI week 4 and week 8, assuming the probit model of Assumption~(\ref{assump1}) holds for values below and above the assay limit. From the tower property of total expectations, 
\begin{equation} \label{P(y=1|M*C)}
    \begin{split}
        P(Y=1|M^*,C) = E(Y|M^*,C) &= E\{ E(Y|M,M^*,C)| M^*,C \} \\
        &= E \{E(Y|M,C)|M^*,C \} ,
    \end{split}
\end{equation}
where the second line follows from Assumption \ref{assump3}.
\begin{theorem}\label{theorem1}
    Under Assumptions~\ref{assump1},~\ref{assump2} and~\ref{assump3},
    \begin{small}
    \begin{equation}
    \begin{split}
    P\left(Y=1\middle| A=0,M^\ast=m^\ast,C=c\right) &= 
    \Phi\left(\frac{\beta_0+\beta_1 \mu_{M|m^\ast,c} +\beta_2c}{\sqrt{1+\beta_1^2 \sigma^2_{M|m^\ast,c}}}\right) \\
   &= \Phi\left(\frac{\beta_0+\beta_1[(1-\lambda)(\alpha_0 + \alpha_1 c) + \lambda m^\ast]+\beta_2c}{\sqrt{1+\beta_1^2 \lambda\sigma_U^2}}\right) 
    , \label{thm3.1}
    \end{split}
\end{equation}
\end{small}
with $\lambda = \sigma_M^2/ \sigma_{M^\ast}^2$ the reliability ratio, $\mu_{M|m^\ast,c} = (1-\lambda)(\alpha_0 + \alpha_1 c) + \lambda m^\ast$ and $\sigma^2_{M|m^\ast,c} = \lambda\sigma_U^2$.
\end{theorem}

\begin{corollary} \label{cor1}
   Under Assumptions~\ref{assump1},~\ref{assump2} and~\ref{assump3}, the probit regression of $Y$ on $(M^\ast, C)$, 
    \begin{equation}
    P\left(Y=1\middle| M^\ast=m^\ast,C=c\right)=\Phi\left(\beta_0^\ast+\beta_1^{\ast}m^\ast+\beta_2^{\ast}c\right), \label{eq4}
\end{equation}
\end{corollary}
is correctly specified with $\beta^\ast \neq \beta$.
\begin{corollary}\label{cor2}
    Under Assumptions~\ref{assump1},~\ref{assump2} and~\ref{assump3}, the true regression coefficients $\beta$ of Assumption~\ref{assump1} can be expressed in terms of the $\beta^\ast$ from Corolloary~\ref{cor1} as
    \begin{align*}
        \beta_1 &=\frac{\beta_1^\ast}{\sqrt{|\lambda^2-\beta_1^{\ast2}\lambda\sigma_U^2}|},
\end{align*}
\begin{align*}
             \beta_0 &=\beta_0^\ast\sqrt{1+\beta_1^2\lambda\sigma_U^2}-\beta_1\left(1-\lambda\right)\alpha_0,
             \end{align*}
\begin{align*}
          \beta_2 &=\beta_2^\ast\sqrt{1+\beta_1^2\lambda\sigma_U^2}-\beta_1\left(1-\lambda\right)\alpha_1.
\end{align*}
\end{corollary}

\noindent
The pure indirect effect or organic indirect effect relative to $a=0$ of a treatment that shifts the mediator distribution to the left by $\xi$ can be written using the Mediation Formula (\cite{Pearl2010})
    
    \begin{align*}
   & {\mathbb{E}}[Y^{(0,I=1)}] - {\mathbb{E}}[Y^{(0)}] \\
   & = \int_{m,c}{\mathbb{E}}[Y|M=m, A=0,C=c]({f}_{M|A=1,C=c}(m)-{f}_{M|A=0,C=c}(m))f_C(c)dmdc \\
    & = \int_{m,c}{\mathbb{E}}[Y|M=m, A=0, C=c]({f}_{M|A=0,C=c}(m+\xi)-{f}_{M|A=0,C=c}(m))f_C(c)dmdc. 
    \end{align*}

\begin{theorem} \label{theorem2}
    Under Assumptions~\ref{assump1},~\ref{assump2},~\ref{assump3} and using the Mediation Formula, for a treatment that shifts the distribution of the viral reservoir on the log$_{10}$ scale by $\xi$ to the left, the pure or organic indirect effect is identified by
    \begin{small}

\begin{align}
   & {\mathbb{E}}[Y^{(0,I=1)}] - {\mathbb{E}}[Y^{(0)}] \hspace{11cm} \notag\\
    & \hspace{0.5cm} = \sum_{c=0}^1 \left[ \Phi\left(\frac{{\beta}_0 + {\beta}_1({\alpha}_0 + {\alpha}_1c -\xi)+{\beta}_2c}{\sqrt{1+({\beta}_1{\sigma}_M)^2}} \right) - \Phi\left(\frac{{\beta}_0 + {\beta}_1({\alpha}_0 + {\alpha}_1c )+{\beta}_2c}{\sqrt{1+({\beta}_1{\sigma}_M)^2}} \right) \right]P(C=c).
    \end{align}

\end{small}

\end{theorem}
The proofs of Theorems ~\ref{theorem1}, ~\ref{theorem2} and Corollary~ \ref{cor2} are provided in Section A of the Supplementary Material (\cite{Herath2025}).
\subsubsection{Estimation Procedure}
The following procedure is used to estimate the pure or organic indirect effect of a new curative treatment $A$, hypothesized to reduce viral persistence measures: shifting their distributions on the log$_{10}$ scale to the left by 0.5log$_{10}$, 1log$_{10}$  (indicating a 10-fold reduction), 1.5log$_{10}$, and 2log$_{10}$ (indicating a 100-fold reduction), given the pre-treatment common cause, whether participants initiated ART during acute/early HIV infection
versus chronic HIV infection.

Let $\delta$ be an indicator representing whether a mediator value $M^\ast$ is above the assay lower limit ($AL$), specifically: $\delta=1$ if $M^\ast > AL$ and $\delta=0$ if $M^\ast \leq AL$. The following algorithm summarizes the estimation procedure: 

\begin{enumerate}
    \item \textbf{Model fitting}: Fit the following generalized linear models on data from participants not receiving curative treatments:
    \begin{align*}
        &\mathbb{E}\left[M_i^\ast \mid A_i=0, C_i\right] = \alpha_0 + \alpha_1C_i, \\
        &\Phi^{-1}\left(\mathbb{E}[Y_i \mid A_i=0, M_i^\ast, C_i]\right) = \beta_0^\ast + \beta_1^\ast M_i^\ast + \beta_2^\ast C_i.
    \end{align*}
    Following the method of \cite{Chernofsky}, who demonstrated that numerical optimization efficiently handles the assay limit problem across different sample sizes and mediator shifts (compared to extrapolation and imputating $AL/2$), we calculate the maximum likelihood estimators for $\alpha$ and $\beta$ using numerical integration and optimization of the observed data likelihood: 
    \vspace{1em}
    \begin{enumerate}
        \setlength\itemsep{1em}
   
    \item[1a.] \textbf{Observed data likelihood:} The observed data likelihood $L$ can be expressed as:
    \begin{scriptsize}
    \begin{align*}
    L = \prod_{i=1}^N \left[f(y_i|m_i^*, c_i; \beta^*)f(m_i^*|c_i;\alpha, \sigma_{M^*}^2) \right]^{\delta_i}\times \left[\int_{-\infty}^{AL}f(y_i|m^*, c_i; \beta^*)f(m^*|c_i; \alpha, \sigma_{M^*}^2)dm^*\right]^{1-\delta_i},
\end{align*}
\end{scriptsize}
where $AL$ represents the assay lower limit on the $log_{10}$ scale.

\item[1b.] \textbf{Log likelihood:} The log-likelihood $l$ is given by:
\begin{footnotesize}
    \begin{align*}
        l &= \sum_{i=1}^N \delta_i \biggr[y_i log \left(\Phi(\beta_0^* + \beta_1^*m_i^* +\beta_2^{*T}c_i)\right)+ (1 - y_i)log (1 - \Phi(\beta_0^* + \beta_1^*m_i^* +\beta_2^{*T}c_i))  \\ & \hspace{1cm}- \frac{1}{2}log(2\pi\sigma_{M^*}^2) -  \frac{(m^* - \alpha^Tc_i)^2}{2\sigma_{M^*}^2}\biggr] 
         + (1- \delta_i) log \int_{-\infty}^{AL}\biggl[  \left(\Phi(\beta_0^* + \beta_1^*m_i^* +\beta_2^{*T}c_i)\right)^{y_i} \\ & \hspace{1cm} * \left(1 - \Phi(\beta_0^* + \beta_1^*m_i^* +\beta_2^{*T}c_i)\right)^{1-y_i} *  \frac{1}{\sqrt{2\pi\sigma_{M^*}^2}}\exp{\left( \frac{-(m^{*} - \alpha^{T}c_{i})^{2}}{2\sigma_{M^{*}}^{2}}\right)}\biggr]dm^*.
    \end{align*}
    \end{footnotesize}
     \end{enumerate}
    \item 	\textbf{Estimate maximum likelihood estimators $\hat{\beta}$} : Obtain the measurement-error-adjusted estimate $\hat{\beta}$ from $\hat{\beta^\ast}$, as detailed in Corollary~\ref{cor2}.\label{est_b}
    \item \textbf{Estimate the pure indirect effect:} Calculate the maximum likelihood estimator for the pure or organic indirect effect using Theorem~\ref{theorem2}.
\end{enumerate}
\noindent
In step 3, the pure indirect effect/organic indirect effect relative to $a=0$ is estimated by plugging in $\hat{\alpha}$ and $\hat{\beta}$ from step \ref{est_b}. This is the maximum likelihood estimate (MLE) of the pure or organic indirect effect of an HIV curative treatment that shifts the distribution of the viral reservoir.

\cite{Chernofsky} proposed an alternative method for estimating the pure indirect effect/organic indirect effect relative to $a=0$ after completing step 1 of the algorithm above. In addition, they did not account for measurement error when estimating the indirect effect. Their algorithm:
\begin{enumerate}
 \setlength\itemsep{1em}
    \item Fit the following generalized linear model on data from participants not receiving curative interventions:
  \begin{align*}
        &\mathbb{E}\left[M_i \mid A_i=0, C_i\right] = \alpha_0 + \alpha_1C_i \\
        &g\left(\mathbb{E}[Y_i \mid A_i=0, M_i, C_i]\right) = \beta_0 + \beta_1 M_i + \beta_2 C_i,
    \end{align*}
    where $g(.)$ is the logit link, likelihood similar to the one in this article, but for a logit link.
    \item Using $\hat{\alpha}$, sample $M_{ij}$ for $j = 1, ..., J $ and for those $i = 1, ..., N $ with $\delta_i = 0 $ from a normal distribution truncated to $M^\ast \leq AL.$
    \item For a treatment that shifts the viral reservoir downwards by $\xi$, predict
     \begin{equation*}
  \hat{\mathbb{E}}[Y_i|A_i=0, M_i-\xi, C_i] =
    \begin{cases}
      g^{-1} (\hat{\beta}_0 + \hat{\beta}_1(M_i - \xi) + \hat{\beta}_2C_i )& \text{if $\delta_i = 1$}\\
      \frac{1}{J}\sum_{j=1}^J  g^{-1} (\hat{\beta}_0 + \hat{\beta}_1(M_{ij} - \xi) + \hat{\beta}_2C_i) & \text{if $\delta_i =0.$}
    \end{cases}  
    \end{equation*}
    \item Estimate the pure indirect effect/ organic indirect effect relative to $a=0$ by 
    \begin{equation*}
        \hat{\mathbb{E}}[Y_i^{(0, I=1)}] - \hat{\mathbb{E}}[Y_i^{(0)}] = \frac{1}{N} \sum_{i=1}^N (\hat{\mathbb{E}}[Y_i \mid M_i = m_i - \xi, A_i=0, C_i=c_i] - y_i).
    \end{equation*}
\end{enumerate}

\section{Results} \label{Results}
Table~\ref{charac} summarizes the baseline characteristics of the 104 participants who were not receiving HIV curative treatments and were not on NNRTIs in the six ACTG analytic treatment interruption (ATI) studies. 

\begin{table}[H]
\caption{Baseline characteristics of AIDS Clinical Trials Group (ACTG) participants included in this study. }
\label{charac}
\begin{tabular}{@{}lc@{}}
\hline
Characteristic &  Total $(N=104)$\\
\hline
Sex, male, $N(\%)$    &   92 (88\%) \\
Age, median years (Q1, Q3)    &  43 (36, 49) \\
Race/ethnicity, $N(\%)$    &   \\
 \hspace{0.5cm}White, non-Hispanic    &  66 (63\%) \\
 \hspace{0.5cm}Black, non-Hispanic   &  14 (13\%)\\
 \hspace{0.5cm}Hispanic   &  14 (13\%) \\
 \hspace{0.5cm}Other    & 10 (10\%) \\
Source study, $N(\%)$    &   \\
 \hspace{0.5cm}$A5170$ &  27 (26\%) \\
 \hspace{0.5cm}$A5197$ &  7 (7\%) \\
 \hspace{0.5cm}$A5068$ & 3 (3\%) \\
 \hspace{0.5cm}$A5024$ & 2 (2\%) \\
 \hspace{0.5cm}$ACTG371$ & 20 (19\%) \\
 \hspace{0.5cm}$A5345$ & 45 (43\%) \\
Chronic HIV infection (vs. acute/early) at ART initiation, $N(\%)$ & 72 (69\%) \\
\hline
\end{tabular}
\end{table}

Of the 104 participants, 72 initiated ART during chronic infection and 32 during acute or early HIV infection. We chose the target population similarly, with 69\% starting ART during chronic infection. Thus, for the confidence intervals, we did not account for variance in the distribution of the pre-treatment common causes. Among the 104 participants, $37\%$ maintained viral suppression through ATI week 4, while only $14\%$ maintained viral suppression through ATI week 8. 

Table A1 on the Supplementary Material (\cite{Herath2025}) presents the fitted models for the mediators and for the outcomes. The standard deviation for the true underlying caRNA and SCA on the log$_{10}$ scale were estimated as 0.79 (95\% CI: 0.52, 1.06) and 0.84 (95\% CI: 0.61, 0.93). The standard deviation of the error for caRNA and SCA were estimated as 0.29 and 0.47 log$_{10}$, respectively, using data from completed ACTG studies (\cite{Scully2022}, \cite{Riddler2016}). 

Table~\ref{ie_results} presents the pure indirect effect or organic indirect effects relative to no treatment of curative HIV treatments that shift the distribution of the HIV persistence measures downward. Table~\ref{ie_results} includes results both adjusted and unadjusted for measurement error, along with 95\% confidence intervals calculated with the Delta method. Notably, the estimated indirect effects adjusted for measurement error were systematically higher and more precise, revealing an underestimation in the unadjusted analysis, consistent with \cite{Valeri2014_ME}.

A one-log$_{10}$ downward shift on the log$_{10}$ scale (equivalent to a 10-fold reduction) in caRNA leads to an indirect effect of a $24.1\%$ (95\% CI: $12.6\%, 35.5\%$) increase in the probability of viral suppression through week 4 after ATI, from 34.6\% to 58.7\%. A two-log$_{10}$ downward shift on the log$_{10}$ scale (100-fold reduction) in caRNA leads to an indirect effect of a $45.2\%$ ($95\%$ CI: $28.6\%, 61.7\%$) increase in the probability of viral suppression through week 4 after ATI, from $34.6\%$ to $79.8\%$. 

In contrast, a one-log$_{10}$ downward shift on the log$_{10}$ scale (10-fold reduction) in SCA leads to an indirect effect of a $10.7\%$ ($95\%$ CI: $-4.7\%, 26.1\%$) increase in the probability of viral suppression by week 4 after ATI, from $36.4\%$ to $47.1\%$. A two-log$_{10}$ downward shift on the log$_{10}$ scale (100-fold reduction) in SCA leads to an indirect effect of a $21.6\%$ ($95\%$ CI: $-8.6\%, 51.9\%$) increase in the probability of viral suppression by week 4 after ATI, from $36.4\%$ to $58.0\%$.

The indirect effect mediated by caRNA on viral suppression through week 8 after ATI is lower compared to week 4 ($17.7\% $ vs. $24.1\%$ for a 10-fold reduction). In contrast, the indirect effect mediated by SCA on viral suppression through week 8 after ATI is higher compared to week 4 ($19.8\%$ vs. $10.7\%$ for a 10-fold reduction).  

Confidence intervals based on the Delta method are presented throughout instead of bootstrap confidence intervals, as the latter caused numerical instability when calculated for week 8 due to the small number of virologically suppressed participants at week 8. The 95\% bootstrap confidence interval (Table A4 on the Supplementary Material (\cite{Herath2025})) for caRNA through week 4 shows that we get similar results for confidence intervals using both the Delta method and nonparametric bootstrap confidence intervals. 

A sensitivity analysis using a logit link instead of a probit link produced approximately the same results for the week 4 analysis (Table A3 of the Supplementary Material \cite{Herath2025}). For the week 8 analysis, the indirect effect found using the logit link was larger than using probit link. Hence, the week 8 analysis appears somewhat more sensitive to model specification.
The assay limit in the data source used for the measurement error SD estimate for caRNA was 49 copies per million CD4+ T cells and for SCA it was 1 copy/ml. 
A sensitivity analysis was also conducted using a single assay limit across all ACTG studies in our analysis (with assay limits of 92 copies per million CD4+ T cells for caRNA and 1 copy/ml for SCA). This change did not qualitatively affect the comparison between caRNA and SCA (Table A2 on the Supplementary Material).

\begin{table}[H]
\caption{Pure indirect effects/organic indirect effects relative to no treatment of curative HIV treatments that shift the distribution of HIV persistence measures on the $log_{10}$ scale downwards.}
\label{ie_results}
\begin{threeparttable}
\begin{tabular}{@{}p{2.4cm} >{\centering\arraybackslash}p{2cm} c >{\centering\arraybackslash}p{2.1cm} >{\centering\arraybackslash}p{2.1cm} >{\centering\arraybackslash}p{2.7cm}@{}}
\toprule
\makecell{HIV Persistence \\ Measure} & \makecell{Mediator \\ Shift\tnote{a}} & Week\tnote{b} & 
\multicolumn{2}{c}{Indirect Effect\tnote{c}} & \makecell{95\% CI\tnote{d}} \\
\cmidrule(lr){4-5}
& & & Measurement Error Ignored & Measurement Error Adjusted & \\
\midrule
 
     caRNA\tnote{e,f} & 0.5 & 4   & 10.4\% & 11.9\% & (6.0\%, 17.8\%)\\
     caRNA            & 1.0 & 4   & 21.2\% & 24.1\% & (12.6\%, 35.5\%)\\
     caRNA            & 1.5 & 4   & 31.5\% & 35.5\% & (20.2\%, 50.7\%)\\
     caRNA            & 2.0 & 4   & 40.8\% & 45.2\% & (28.6\%, 61.7\%)\\ \hline
     SCA\tnote{g,h} & 0.5 & 4  & 4.0\% & 5.3\%  & (-2.3\%, 12.8\%)\\
     SCA            & 1.0 & 4  & 8.1\% & 10.7\% & (-4.7\%, 26.1\%)\\
     SCA            & 1.5 & 4  & 12.2\% & 16.2\% & (-6.9\%, 39.3\%)\\
     SCA            & 2.0 & 4  & 16.4\% & 21.6\% & (-8.6\%, 51.9\%)\\ \hline
     caRNA\tnote{e,f} & 0.5 & 8  & 6.6\% & 7.7\% & (2.1\%, 13.2\%)\\
     caRNA            & 1.0 & 8  & 15.1\% & 17.7\% & (4.4\%, 31.0\%)\\
     caRNA            & 1.5 & 8  & 25.2\% & 29.6\% & (7.7\%, 51.4\%)\\
     caRNA            & 2.0 & 8  & 36.2\% & 42.3\% & (13.2\%, 71.4\%)\\ \hline
     SCA\tnote{g,h}  & 0.5 & 8   & 6.1\% & 8.4\%  & (1.4\%, 15.3\%)\\
     SCA             & 1.0 & 8   & 14.0\% & 19.8\%  & (2.6\%, 36.9\%)\\
     SCA             & 1.5 & 8   & 23.5\% & 33.4\%  & (5.2\%, 61.7\%)\\
     SCA             & 2.0 & 8   & 34.1\% & 47.8\%  & (11.1\%, 84.6\%) \\ \hline
    \end{tabular}
        \begin{footnotesize}
    The estimated probability of virologic suppression through ATI week 4 without curative treatment was 36/104 (or 34.6\%) and 13/104 (or 12.5\%) through ATI week 8 in 104 participants with caRNA measured. \\
     The estimated probability of virologic suppression through ATI week 4 without curative treatment was 32/88 (or 36.4\%) and 10/88 (or 11.4\%) through ATI week 8 in 88 participants with SCA measured. 
    
    \tnote{a} Curative treatment-induced downward shift of the viral persistence measure distribution, given the common cause, on the $log_{10}$ scale (compared to no curative treatment).
     
     \tnote{b} Viral rebound in the first 4 weeks and first 8 weeks after ART interruption. 
     
    \tnote{c} Pure or organic indirect effect relative to no treatment: the difference in probability of virologic suppression that is mediated through the HIV persistence measure.
    
    \tnote{d} 95\% Confidence Interval (measurement error adjusted) calculated using the Delta method.
    
    \tnote{e} Cell-associated HIV RNA, on ART. All 104 participants had caRNA measured. Assay limit of ACTG A5345: 28 copies per million CD4+ T cells. Assay limit of remaining studies: 92 copies per million CD4+ T cells.
    
    \tnote{f} $log_{10}$ caRNA: estimated measurement error SD = 0.29 (\cite{Scully2022}).
    
    \tnote{g} Single-copy plasma HIV RNA, on ART. Restricted to the 88 participants with SCA measured. Assay limit of ACTG A5345: 0.56 copies per ml. Assay limit of remaining studies: 1 copy per ml.
    
    \tnote{h} $log_{10}$ SCA: estimated measurement error SD = 0.47 (\cite{Riddler2016}).
    \end{footnotesize}
\end{threeparttable}

\end{table}

\section{Simulation Results} \label{Sim_section}
 The simulation study is designed based on the HIV curative treatment application described in Section~\ref{Results}, which aims to estimate the pure indirect effect/organic indirect effect relative to no treatment of a curative HIV treatment that shifts the distribution of HIV persistence measures on the log$_{10}$ scale downwards. The simulation study assesses the performance of the estimators under various shifts on the distribution of the log$_{10}$-transformed mediator for different sample sizes. The simulated data closely follow the estimated distribution from the HIV curative treatment analysis presented in Section~\ref{Results} and the models described in Section~\ref{stat_analysis}. 
 
 The simulations focus on both viral persistence measures, cell-associated HIV RNA, and single-copy plasma HIV RNA, and on viral suppression through week 4. In the simulations, under $a=0$, caRNA, the log$_{10}$ viral reservoir, follows a normal distribution with mean $\alpha_0 + \alpha_1c$ and variance $\sigma_M^2$ with $(\alpha_0, \alpha_1) = (1.57, 0.88)$ and $\sigma_M = 0.58$. The measurement error follows a normal distribution with mean $0$ and variance $\sigma_U = 0.29$. The distribution of the pre-treatment common cause is fixed, where the percentage of chronically HIV-infected participants at ART initiation fixed at 69\%. The distribution of $Y$, viral suppression through week 4, is Bernoulli with probability of $Y=1$ equal to $p_{\beta}(m,c) = \Phi(\beta_0 + \beta_1m + \beta_2c)$ with $(\beta_0, \beta_1, \beta_2) = (1.36, -1.11, 0.84)$.

  Under $a=0$, SCA, the $log_{10}$ viral reservoir follows a normal distribution with mean $\alpha_0 + \alpha_1c$ and variance $\sigma_M^2$ with $(\alpha_0, \alpha_1) = (-0.02, 0.10)$ and $\sigma_M = 0.65$. The measurement error follows a normal distribution with mean $0$ and variance $\sigma_U = 0.47$.  The distribution of the pre-treatment common cause is fixed, where the percentage of chronically HIV-infected participants at ART initiation fixed at 72\%. The distribution of $Y$, viral suppression through week 4, is Bernoulli with probability of $Y=1$ equal to $p_{\beta}(m,c) = \Phi(\beta_0 + \beta_1m + \beta_2c)$ with $(\beta_0, \beta_1, \beta_2) = (-0.22, -0.34, -0.16)$. 

 The simulated scenarios evaluate the effect of different sample sizes and mediator shifts on the estimators. A sample of $N=104$ for caRNA and $N=88$ for SCA resembles our data example, while a larger sample size of $N=1000$ reflects larger sample properties.

\begin{figure}[H]
\includegraphics[scale = 0.65]{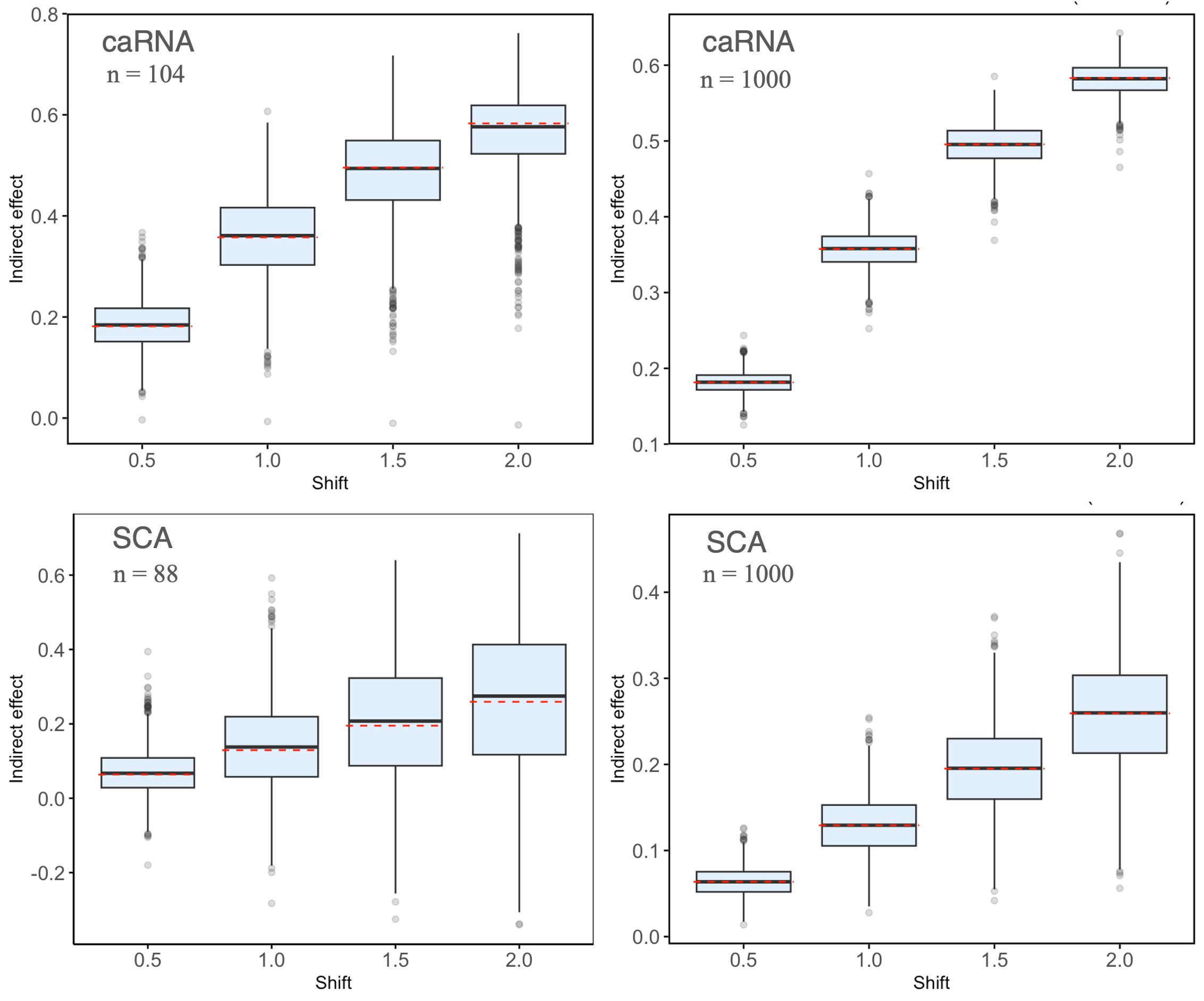}
\caption{Simulation study results showing estimator performance under varying shifts of log-transformed mediators with measurement error (caRNA and SCA). The top row presents simulated estimates for caRNA with $N = 104$ and $N = 1000$; the bottom row for SCA with $N = 88$ and $N = 1000$. Mediator shifts are 0.5, 1.0, 1.5, and 2.0. True indirect effects are shown as dashed lines. }
\label{Si_boxplots}
\end{figure}

 Table~\ref{sim_results_ME} presents the true indirect effect, bias, root mean square error (rMSE), and coverage probability based on 2000 simulated datasets, covering both scenarios where measurement error in the mediators was ignored and where it was incorporated.  
Figure~\ref{Si_boxplots} and Table~\ref{sim_results_ME} shows the influence of sample size and different mediator shifts on the performance of the estimators when the measurement error is incorporated. Notably, for the mediator SCA under a two-log$_{10}$ downward shift on the log$_{10}$ scale, the estimates show instability, potentially due to extrapolation beyond the support of the data and the smaller sample size compared to caRNA. 
In contrast, when measurement error is ignored, the results show substantial bias and low coverage probabilities for both mediators across different shifts.

\begin{landscape}
\begin{table}[H]
\caption{Results of the simulation study evaluating the method used to estimate the pure or organic indirect effects of curative HIV treatments that shift the distribution of HIV persistence measures downwards, with and without incorporating measurement error.}
\label{sim_results_ME}
\begin{threeparttable}
 \begin{tabular}{@{}l >{\centering\arraybackslash}p{3cm} >{\centering\arraybackslash}p{2cm} c 
                >{\centering\arraybackslash}p{1.9cm} c >{\centering\arraybackslash}p{2.5cm} c c >{\centering\arraybackslash}p{2.5cm} c@{}}
    \toprule
    & & & & \multicolumn{3}{c}{Measurement Error Ignored} 
      & \multicolumn{3}{c}{Measurement Error Adjusted} \\[-1ex]
    \cmidrule(lr){5-7} \cmidrule(lr){8-10} 
    \makecell{Sample\\ Size} & \makecell{HIV Persistence \\ Measure} 
    & \makecell{Mediator \\ Shift\tnote{a}} & \makecell{True Indirect \\ Effect\tnote{b}} 
    & Bias & rMSE & \makecell{Coverage \\ Probability} 
    & Bias & rMSE & \makecell{Coverage \\ Probability} \\ 
  \hline
    104 &  caRNA & 0.5 & 18.2\% & -3.6\% & 5.1\% & 85.0\% & 0.3\% & 5.0\% &  92.8\%\\
        &        & 1.0 & 35.7\% & -6.8\% & 9.6\% & 87.2\% & 0.0\% & 8.4\% & 93.3\% \\
        &       & 1.5 & 49.6\% & -8.3\% & 11.9\% & 90.9\% & -1.0\% & 9.1\% & 95.1\% \\
        &     & 2.0 & 58.3\% & -7.9\% & 11.7\% & 94.9\% & -1.9\% & 8.3\% &  96.3\% \\ \hline
   88 &   SCA & 0.5 & 6.4\% & -2.1\% & 4.2\% &  91.3\% & 0.6\% & 6.5\% &  97.1\% \\
      &          & 1.0 & 12.9\% & -4.3\% & 8.4\% & 90.6\% & 1.1\% & 12.2\% &  94.4\% \\
      &           & 1.5 & 19.5\% & -6.6\% & 12.5\% & 90.3\% & 0.8\% & 16.8\% &  90.8\% \\
      &          & 2.0 & 25.9\% & -8.9\% & 16.4\% & 89.9\% & -0.2\% & 20.2\% &  86.9\% \\ \hline
     1000 & caRNA & 0.5 & 18.2\% & -3.6\% & 3.8\% & 9.7\%& 0.0\% & 1.5\% & 94.4\% \\
          &      & 1.0 & 35.7\% & -6.8\% & 7.0\% & 10.3\% & 0.0\% & 2.5\% &  94.6\% \\
          &     & 1.5 & 49.6\% & -7.7\% & 8.1\% & 14.1\% & -0.1\% & 2.7\% &  94.7\% \\
          &    & 2.0 &  58.3\% & -6.6\% & 7.1\% & 24.3\% & -0.2\% & 2.2\% &  95.3\% \\ \hline
     1000 & SCA  & 0.5 & 6.4\% & -2.2\% & 2.5\% & 47.4\% & 0.0\% & 1.7\% &  95.0\% \\
           &      & 1.0 & 12.9\% & -4.5\% & 5.0\% & 47.6\% & 0.0\% & 3.4\% &  94.9\% \\
           &     & 1.5 & 19.5\% & -6.8\% & 7.6\% & 48.2\% & 0.0\% & 5.0\% &  94.5\% \\
           &    & 2.0 & 25.9\% & -8.9\% & 9.9\% & 49.6\% & 0.0\% & 6.4\% &  94.0\% \\ \hline
    \end{tabular}
    \label{tab:sim_table}
    \begin{footnotesize}

\tnote{a}Curative treatment-induced downward shift of the viral persistence measure distribution, given the common cause, on the $log_{10}$ scale (compared to no curative treatment).

\tnote{b} The calculated true pure or organic indirect effect. 

Abbreviations: rMSE: root mean square error, bias: average bias, coverage probability: the probability that the 95\% confidence interval captures the true value. \\ For each of the simulated 2000 datasets, a 95\% confidence interval is estimated using the Delta method.
    \end{footnotesize}
\end{threeparttable}
\end{table}
\end{landscape}

\section{Discussion}\label{Discussion}
We estimated the pure indirect effect/organic indirect effect of a curative HIV intervention that shifts the distribution of two viral persistence measures~- cell-associated HIV RNA and single-copy plasma HIV RNA- given pre-treatment common causes. These indirect effects inform the reduction in the viral reservoir necessary to meaningfully increase the probability of viral suppression through week 4 and week 8. Additional estimation tools are required when the mediator is subjected to an assay limit and is measured with error. Our indirect effect estimates build on the work of \cite{Chernofsky}, who addressed the assay limit problem by assuming that the distribution of the outcome depends on how far the mediator values fall below the assay limit. To accommodate measurement error in the mediators, we modeled the binary outcome with a probit regression model, extending the model to mediator values below the assay limit and adjusted for measurement error. Although the sample size in our study is relatively small, it is a large sample for analytic treatment interruption (ATI) studies. 

The findings in Table~\ref{ie_results} highlight the importance of accounting for measurement error in the mediator when estimating indirect effects. For the outcome viral suppression through week 8, adjusting for measurement error increased the magnitude of the indirect effects for caRNA and SCA by approximately 1.2-fold and 1.5-fold, respectively, consistent with the larger measurement error SD for SCA. Without adjustment, measurement error in a mediator introduces bias, leading to underestimation of indirect effects and potentially misleading conclusions. In the application, adjusting for measurement error increased the indirect effect estimates. In the simulations, the confidence intervals showed much better coverage probabilities. Thus, adjusting for measurement error resulted in more accurate and reliable findings that better reflect the underlying causal relationships.

A 10-fold reduction in caRNA resulted in a pure or organic indirect effect of a $24.1\%$ ($95\%$ CI: $12.6\%, 35.5\%$) increase in the probability of viral suppression through week 4 after ATI, while an equivalent shift in SCA led to a pure or organic indirect effect increase of $10.7\%$ ($95\%$ CI: $-4.7\%, 26.1\%$). This suggests that to improve short term virologic suppression, targeting caRNA may be a more effective treatment strategy for HIV curative treatments than targeting SCA at week 4. While the week 4 results are more precise and reliable, the week 8 results may be more important from the perspective of HIV curative treatment.
However, we should note that, based on the simulations, the indirect effect estimate mediated by SCA becomes unreliable after a 10-fold reduction due to sample size limitations leading to data extrapolation.

The assay limits for both caRNA and SCA are not consistent across all ACTG studies. HIV persistence/reservoir assays continue to be refined and optimized. The assay limits used in the primary analysis reflect the operative limits applicable to the specific dataset and testing platform, so the results can vary depending on which assay limits are applied. In the sensitivity analysis using a single, consistent assay limit across all ACTG studies for each viral persistence measure, a 10-fold reduction in caRNA resulted in a pure or organic indirect effect of a $35.8\%$ ($95\%$ CI: $19.4\%, 52.4\%$) increase in the probability of viral suppression through week 4 after ATI, while an equivalent shift in SCA led to a pure or organic indirect effect of increase $12.9\%$ ($95\%$ CI: $-9.4\%, 35.2\%$) (Table A2 on the Supplementary Material). 

Our results show that a substantial reduction of the viral reservoir is necessary to meaningfully reduce the risk of viral rebound through week 4 after ATI. Our findings align with \cite{Hill2014}, who used mathematical modeling and suggested a 2000-fold reduction of the HIV reservoir is necessary for one year without viral rebound. This would be a downward shift of approximately 3.3-log$_{10}$ on the log$_{10}$ scale in Table~\ref{ie_results}. 

Our work on indirect effect estimation on viral suppression builds on \cite{Chernofsky}, who demonstrated a reduction in the risk of viral rebound for various mediator shifts. However, their study did not address measurement error in the mediators, and did not analyze virologic suppression through week 8.  The differences between our findings and theirs are due to the consideration of measurement error and the difference in populations, which provides us with improved inference and aligns with current HIV cure trials which exclude individuals on NNRTIs (\cite{Gunst2025}), in contrast to the data used in \cite{Lok2021} and \cite{Chernofsky}. The methodological improvements including complete maximum likelihood estimation, consideration of measurement error, and the choice of the target population make the results more useful for HIV curative drug development practice. 

In conclusion, estimating pure indirect effects or organic indirect effects relative to no treatment is possible even in the absence of outcome data under treatment. The indirect effect of HIV curative treatments targeting caRNA on week 4 viral suppression are larger than those targeting SCA, and the indirect effects on week 8 viral suppression are similar. Based on the available data, one could recommend prioritizing the development of HIV curative treatments targeting caRNA over those targeting SCA. However, larger ATI studies without curative HIV treatments are needed to make more informed recommendations.  This article demonstrates how causal mediation analysis can be used to identify potential new treatments with promising indirect effects that warrant further evaluation in randomized clinical trials to estimate their overall effect.

\begin{acks}[Acknowledgments]
 The authors would like to thank the study participants who volunteered for the ACTG ART interruption trials, Dr. J.Z. Li (Brigham and Women Hospital, Boston) and the ACTG sites and teams of the ART interruption studies for the data collection. 
\end{acks}

\begin{funding}
Ronald J Bosch was supported in part by NIH/NIAID award UM1AI068634. and the contributing ACTG studies were supported by NIH/NIAID awards UM1AI068636 and UM1AI106701.
Judith J Lok was supported by the NSF under award DMS 1854934.
\end{funding}

\begin{supplement}
\stitle{Supplement to "Causal indirect effect of an HIV curative treatment: mediators subject to an assay limit and measurement error" }
\sdescription{The web appendix contains additional derivations referenced in sections 3 and 4, sensitivity analysis results using difference assay limits, logit link, 95\% bootstrap confidence intervals through week 4, and additional results referenced in sections 4 and 6.}
\end{supplement}
\begin{supplement}

\stitle{Method implementation in R}

\sdescription{R-code for implementing the estimation methods. The code is also available on GitHub (\url{https://github.com/Vindyh/Indirect-Effect_Mediator_Measurement-Error}). Data are available upon request from \url{sdac.data@sdac.harvard.edu}.} 

\end{supplement}

\newpage

\bibliographystyle{imsart-nameyear.bst} 
\bibliography{ref.bib}       

\newpage

\appendix

\input{supplement}

\end{document}

%% file: supplement.tex
\startlocaldefs
\numberwithin{equation}{section}
\theoremstyle{plain}

\endlocaldefs

\title{Supplement to "Indirect effect on viral suppression of an HIV curative treatment that reduces viral persistence: mediators subject to an assay limit and measurement error" }

 \setcounter{table}{0}
 \renewcommand{\thetable}{A\arabic{table}}
  \section{THEORY} \label{Appendix_A} 

   Appendix~\ref{Appendix_A} provides the proofs for the theorems and corollaries in \textit{Section 3.2}
 
\begin{proof}(\textit{Theorem 3.1})\\
First we derive the distribution of $M$ given $M^\ast$ and $c$. For fixed $C=c$, 
$$\left( \begin{array}{cc}
    M \\
     U
\end{array}\right) \sim \mathcal{N} \left (\left(  \begin{array}{ccc}
\alpha_0 + \alpha_1c\\
0\end{array} \right) , \left(  \begin{array}{ccc}
    \sigma_M^2 & 0 \\
    0 & \sigma_U^2
\end{array}\right)\right).$$
Since $Cov(M,M+U) = Cov(M,M) + Cov(M,U)$ and $Cov(M,U) = 0,$
$$ \left( \begin{array}{cc}
    M \\
     M\ast
\end{array}\right)=\left( \begin{array}{cc}
    M \\
     M+U
\end{array}\right) \sim \mathcal{N} \left (\left(  \begin{array}{ccc}
\alpha_0 + \alpha_1c\\
\alpha_0 + \alpha_1c \end{array} \right) , \left(  \begin{array}{ccc}
    \sigma_M^2  & \sigma_M^2 \\
    \sigma_M^2 & \sigma_M^2 + \sigma_U^2
\end{array}\right)\right).$$
\vspace{0.5cm}
\\
Using properties of the bivariate normal distribution \cite[p.~215-216]{Schinazi2022} with \\ $\lambda = \frac{\sigma_M^2}{\sigma_{M^*}^2} = \frac{\sigma_M^2}{(\sigma_M^2 + \sigma_U^2)} , M|m\ast, c$ is normally distributed with mean 
\begin{equation} \label{conditional mean proof}
    \begin{split}
        \mu_{M|m^*,c} &= \mathbb{E}[M|M + U = m^*,c] \\
        &= \mu_M + \frac{\sigma_{M}^2}{\sigma_{M^*}^2}(m^* - \mu_{M^*}) \\
        &= (\alpha_0 + \alpha_1c) + \lambda(m^* - (\alpha_0 + \alpha_1c)) \\
        &= (1-\lambda)(\alpha_0 + \alpha_1c) + \lambda m^*
    \end{split}
\end{equation}
and variance
\begin{equation} \label{conditional variance proof}
    \begin{split}
        \sigma_{M|M^*,C}^2 &= \sigma_{M}^2 - \sigma_{M}^2(\sigma_{M^*}^2)^{-1}\sigma_{M}^2 \\
        &= \sigma_M^2 \left( 1 - \frac{\sigma_M^2}{\sigma_M^2 + \sigma_U^2}\right ) \\
        &= \sigma_M^2 \left ( \frac{\sigma_U^2}{\sigma_M^2 + \sigma_U^2}\right) \\
        &= \lambda \sigma_U^2.
    \end{split}
\end{equation}

\cite{DBOwen} [equation 10,010.8] states the following result on probit-normal integrals:
\begin{equation} \label{probit-normal integral}
   \int_{-\infty}^{+\infty} \Phi(\mu + x\sigma)\phi(x)dx = \Phi\left( \frac{\mu}{\sqrt{1+\sigma^2}}\right).
\end{equation}
Using equation (3.5) and in the last line equation (\ref{probit-normal integral}),
\begin{equation*} \label{Proof:Probit-normal integral}
    \begin{split}
        & P(Y=1|M^*=m^*, C=c) \hspace{10cm}\\
        & \hspace{1cm} = \int P(Y=1|M=m, C=c) f_{M|M^*,C}(m|M^*=m^*,C=c)dm \\ 
        & \hspace{1cm} = \int \Phi (\beta_0 + \beta_1m + \beta_2 c) \phi \left( \frac{m - \mu_{M|m^*,c}}{\sqrt{\sigma_{M|m^*,c}^2}}\right)\frac{1}{\sqrt{\sigma_{M|m^*,c}^2}}dm \\
        & \hspace{1.5cm}\text{ Let $x = \frac{m - \mu_{M|m^*,C}}{\sqrt{\sigma_{M|m^*,c}^2}}$, then $m = x\sqrt{\sigma_{M|m^*,c}^2} + \mu_{M|m^*,c} $,
 } \\
         & \hspace{1cm} = \int \Phi(\beta_0 + \beta_1(x\sqrt{\sigma_{M|m^*,c}^2} + \mu_{M|m^*,C}) + \beta_2c) \phi(x) \frac{1}{\sqrt{\sigma_{M|m^*,c}^2}}\sqrt{\sigma_{M|m^*,c}^2}dx \\
         & \hspace{1cm} = \int \Phi (\beta_0 + \beta_1 \mu_{M|m^*,c} + \beta_2c + x\beta_1 \sqrt{\sigma_{M|m^*,c}^2}) \phi(x) dx \\
         & \hspace{1cm} = \Phi \left ( \frac{\beta_0 + \beta_1 \mu_{M|m^*,c} + \beta_2c}{\sqrt{1+ \beta_1^2 \sigma_{M|m^*,c}^2}} \right ).
    \end{split}
\end{equation*}
\end{proof}

\begin{proof} (\textit{Corollary 2}) \\
    Equation (3.6) can be written using equations (\ref{conditional mean proof}) and (\ref{conditional variance proof}) as follows:
\begin{equation} \label{P(Y|M*)eq.2}
    \begin{split}
        P(Y=1|M^*=m^*,C=c) &= \Phi \left(\frac{\beta_0 + \beta_1((1-\lambda)(\alpha_0 + \alpha_1c) + \lambda m^*) +\beta_2c}{\sqrt{1+\beta_1^2 \lambda \sigma_U^2 }} \right).
    \end{split}
\end{equation}
Let \begin{equation} \label{r_*}
    r_* = \frac{1}{\sqrt{1+\beta_1^2 \lambda \sigma_{U}^2}}.
\end{equation}
Then using equations (\ref{P(Y|M*)eq.2}) and (\ref{r_*}),
\begin{equation} \label{beta^*}
\begin{split}
    \beta_0^* &= r_*(\beta_0 + \beta_1(1-\lambda)\alpha_0), \\
    \beta_1^* &= r_* \lambda \beta_1, \\
    \beta_2^* &= r_*(\beta_1(1-\lambda)\alpha_1 + \beta_2).
\end{split}
\end{equation}

Considering $\beta_1^*$,
\begin{equation*} \label{eq14}
       \beta_1^* {} = {} \frac{\lambda \beta_1}{\sqrt{1+\beta_1^2 \lambda \sigma_{U}^2 }} 
      \end{equation*}

      \begin{equation*}
    \begin{split}
     & \Rightarrow  \hspace{1cm } \beta_1^{*2} = \frac{\lambda^2 \beta_1^2}{1+\beta_1^2 \lambda \sigma_{U}^2} \hspace{2cm}\\
      & \Rightarrow \hspace{1cm} \beta_1^{*2} (1+\beta_1^2 \lambda \sigma_{U}^2) = \lambda^2\beta_1^2 \\
       & \Rightarrow \hspace{1cm} \beta_1^2( \lambda^2 - \beta_1^{*2} \lambda \sigma_U^2 ) = \beta_1^{*2}\\ 
        & \Rightarrow \hspace{1cm} \beta_1^2 = \frac{\beta_1^{*2}}{\lambda^2 - \beta_1^{*2}\lambda \sigma_{U}^2}. \\
            \end{split}
\end{equation*}

Then, 
\begin{equation*}
\begin{split}    
        \beta_1 &= \frac{\beta_1^*}{\sqrt{|\lambda^2 -\beta_1^{*2}\lambda \sigma_{U}^2 |}},\\
    \end{split}
\end{equation*}
because the signs of $\beta_1$ and $\beta_1^*$ are the same. \\
Then from equation (\ref{beta^*}), $\beta_0$ and $\beta_2$ are given by
\begin{equation*} 
    \begin{split}
        \beta_0 &= \beta_0^* \sqrt{1+\beta_1^2 \lambda \sigma_{U}^2} - \beta_1(1-\lambda)\alpha_0, \\
        \beta_2 &= \beta_2^* \sqrt{1+\beta_1^2 \lambda \sigma_{U}^2} - \beta_1(1-\lambda)\alpha_1.
    \end{split}
\end{equation*}
\end{proof}

\begin{proof} (\textit{Theorem 3.2}) \\
    Recall that the treatment shifts the distribution of the viral reservoir to the left by $\xi$. Then the pure indirect effect or organic indirect effect relative to $a=0$ can be written using the Mediation Formula (equation (3.3))
    
    \begin{align*}
   & {\mathbb{E}}[Y^{(0,I=1)}] - {\mathbb{E}}[Y^{(0)}] \\
   & = \int_{m,c}{\mathbb{E}}[Y|M=m, A=0,C=c]({f}_{M|A=1,C=c}(m)-{f}_{M|A=0,C=c}(m))f_C(c)dmdc \\
    & = \int_{m,c}{\mathbb{E}}[Y|M=m, A=0, C=c]({f}_{M|A=0,C=c}(m+\xi)-{f}_{M|A=0,C=c}(m))f_C(c)dmdc, 
    \end{align*}
    where the second equality follows from a change of variable. \vspace{3cm}
    
    \newpage
    Next, consider $c=0 $ first.
    
    \begin{equation*}
    \begin{split}
      &\int_{m}\mathbb{E}[Y|M=m, A=0, C=0]f_{M|A=0,C=0}(m+\xi)dm \\ 
      & \hspace{1cm} = \int_m\mathbb{E}[Y|M=\Tilde{m}-\xi, A=0, C=0](f_{M|A=0,C=0}(\Tilde{m})d\Tilde{m} \hspace{4cm}\\
        & \hspace{1cm}= \int_m\mathbb{E}[Y|M=\Tilde{m}-\xi, A=0, C=0] \phi \left( \frac{\Tilde{m}-\alpha_0}{{\sigma}_M}\right)\frac{1}{\sigma_M}d\Tilde{m}\\
        & \hspace{1cm}= \int_m g^{-1}(\beta_0 + \beta_1(\Tilde{m} - \xi))\phi\left( \frac{\Tilde{m}-\alpha_0}{\sigma_M}\right)\frac{1}{\sigma_M}d\Tilde{m} \\
         & \hspace{1cm} = \int_m \Phi({\beta}_0 + {\beta}_1(\Tilde{m} - \xi))\phi\left( \frac{\Tilde{m}-{\alpha}_0}{{\sigma}_M}\right)\frac{1}{{\sigma}_M}d\Tilde{m} \\
        & \hspace{1cm} = \int_m \Phi[{\beta}_0 + {\beta}_1( {\alpha}_0 + {\sigma}_M x - \xi)]\phi\left( x\right)dx \\
        & \hspace{1cm} = \int_m \Phi[{\beta}_0 + {\beta}_1( {\alpha}_0  - \xi) + x{\beta}_1{\sigma}_M ]\phi\left( x\right)dx \\
        & \hspace{1cm} = \Phi\left(\frac{{\beta}_0 + {\beta}_1({\alpha}_0-\xi)}{\sqrt{1+({\beta}_1{\sigma}_M)^2}} \right),
    \end{split}
    \end{equation*}
    where in the first equality, $\Tilde{m} = m+\xi$, in the fifth equality, $x = \left (\frac{\Tilde{m}-{\alpha}_0}{{\sigma}_M}\right),$ so that $ \Tilde{m} = {\alpha}_0 + {\sigma}_M x $, and using equation (\ref{probit-normal integral}) in the last line. 
    
    Similarly, using the above with the shift of the viral reservoir $\xi =0:$
    \begin{align*}
             \int_{m}{\mathbb{E}}[Y|M=m, A=0, C=0]{f}_{M|A=0,C=1}(m))f_C(c)dm &= \Phi\left(\frac{{\beta}_0 + {\beta}_1{\alpha}_0}{\sqrt{1+({\beta}_1{\sigma}_M)^2}} \right). 
    \end{align*}
    \newpage
    Next, consider $c=1$ .
    \begin{equation*}
        \begin{split}
            & \int_{m}\mathbb{E}[Y|M=m, A=0, C=1]f_{M|A=0,C=1}(m+\xi)dm \hspace{5cm}\\ 
             & \hspace{1cm} = \int_m\mathbb{E}[Y|M=\Tilde{m}-\xi, A=0, C=1](f_{M|A=0,C=1}(\Tilde{m})d\Tilde{m} \\
              & \hspace{1cm} = \int_m\mathbb{E}[Y|M=\Tilde{m}-\xi, A=0, C=1] \phi \left( \frac{\Tilde{m}-(\alpha_0+\alpha_1)}{{\sigma}_M}\right)\frac{1}{\sigma_M}d\Tilde{m}\\
              & \hspace{1cm}  = \int_m g^{-1}(\beta_0 + \beta_1(\Tilde{m} - \xi) + \beta_2)\phi\left( \frac{\Tilde{m}-(\alpha_0+\alpha_1)}{\sigma_M}\right)\frac{1}{\sigma_M}d\Tilde{m} \\
              & \hspace{1cm} = \int_m \Phi({\beta}_0 + {\beta}_1(\Tilde{m} - \xi)+\beta_2)\phi\left( \frac{\Tilde{m}-(\alpha_0+\alpha_1)}{{\sigma}_M}\right)\frac{1}{{\sigma}_M}d\Tilde{m} \\
            & \hspace{1cm} = \int_m \Phi[{\beta}_0 + {\beta}_1( {\alpha}_0 +\alpha_1 + {\sigma}_M x - \xi)+\beta_2]\phi\left( x\right)dx \\
            & \hspace{1cm} = \int_m \Phi[{\beta}_0 + {\beta}_1( {\alpha}_0  +\alpha_1 - \xi) + \beta_2 + x{\beta}_1{\sigma}_M ]\phi\left( x\right)dx \\
            & \hspace{1cm}  = \Phi\left(\frac{{\beta}_0 + {\beta}_1({\alpha}_0 + {\alpha}_1 -\xi)+{\beta}_2}{\sqrt{1+({\beta}_1{\sigma}_M)^2}} \right),
            \end{split}
    \end{equation*}
        where in the first line, $\Tilde{m} = m+\xi$, in the fifth line, $x = \left (\frac{\Tilde{m}-(\alpha_0+\alpha_1)}{{\sigma}_M}\right),$  so that $  \Tilde{m} = \alpha_0 +\alpha_1 + {\sigma}_M x $, and using equation (\ref{probit-normal integral}) in the last line. 
        
         Similarly, using the above with the shift of the viral reservoir $\xi =0:$
         \begin{align*}
             \int_{m}{\mathbb{E}}[Y|M=m, A=0, C=1]{f}_{M|A=0,C=1}(m))f_C(c)dm &= \Phi\left(\frac{{\beta}_0 + {\beta}_1({\alpha}_0 + {\alpha}_1)+{\beta}_2}{\sqrt{1+({\beta}_1{\sigma}_M)^2}} \right).
         \end{align*}
    
\end{proof}

\newpage
\noindent
\textbf{Variance estimation using the Delta method}
\hspace{1cm}

To estimate the variance of the indirect effect, first consider $ {{\mathbb{E}}}[Y^{(0, I=1)}] $ as a function of the key parameters:
\begin{equation*}
    {{\mathbb{E}}}[Y^{(0, I=1)}] = g({\beta^*}, {\alpha^*}, {\sigma}_{M^*}^2).
\end{equation*}
The derivative of this function with respect to the model parameters is given by: 
\begin{equation*}
    g'({\beta^*}, {\alpha^*}, {\sigma}_{M^*}^2) = \begin{bmatrix}
\frac{\partial g(\beta^*, \alpha^\ast, \sigma^2_{M^\ast})}{\partial \beta^*} & \frac{\partial g(\beta^*, \alpha^\ast, \sigma^2_{M^\ast})}{\partial \alpha^*} & \frac{\partial g(\beta^*, \alpha^\ast, \sigma^2_{M^\ast})}{\partial \sigma_{M^*}^2}  
\end{bmatrix} .
\end{equation*}
For variance estimation, recall that the variance of the parameters is approximated by:
\begin{equation*}
    {\text{V}}_{\beta^*, \alpha^*, \sigma_{M^*}^2} = {{VAR}}({\beta^*}, {\alpha^*}, {\sigma}_{M^*}^2) \approx \frac{1}{nI({\beta^*}, {\alpha^*}, {\sigma}_{M^*}^2)},
\end{equation*}
where $ I({\beta^*}, {\alpha^*}, {\sigma}_{M^*}^2)$ is the Fisher information for one observation.

Using the Delta method, the variance of the pure indirect effect or organic indirect effect relative to $a=0$ can be estimated as 
\begin{equation*}
    \widehat{VAR}(\text{Indirect effect}) = g'(\widehat{\beta}^*, \widehat{\alpha}^*, \widehat{\sigma}_{M^*}^2) . \widehat{\text{V}} _{\beta^*, \alpha^*, \sigma_{M^*}^2} . g'(\widehat{\beta}^*, \widehat{\alpha}^*, \widehat{\sigma}_{M^*}^2)^T \end{equation*}
This provides an approximation for the uncertainty in the estimated indirect effect, accounting for variability in the parameter estimates.

The following proof shows the derivatives for $ {{\mathbb{E}}}[Y^{(0, I=1)}] $ in equation~(3.8). Similar derivatives for $\mathbb{E}[Y^{(0)}]$ in equation~(3.8) can be derived with the shift of the viral reservoir $\xi = 0$.\\
To prepare for taking derivatives of the function $g$, first recall that
\begin{equation*}
        \begin{split}
        \lambda &= \frac{\sigma_M^2}{\sigma_{M^*}^2} = \frac{\sigma_M^2}{(\sigma_M^2 + \sigma_U^2)} , \\
        \sigma_U^2 &= \sigma_{M^*}^2(1-\lambda),
    \end{split}
\end{equation*}
\begin{align}
        \beta_1 &= \frac{\beta_1^*}{\sqrt{ |\lambda^2 -\beta_1^{*2}\lambda \sigma_{U}^2 }|}, \label{beta1} \\
        \beta_0 &= \beta_0^* \sqrt{1+\beta_1^2 \lambda \sigma_{U}^2} - \beta_1(1-\lambda)\alpha_0, \notag \\
        \beta_2 &= \beta_2^* \sqrt{1+\beta_1^2 \lambda \sigma_{U}^2} - \beta_1(1-\lambda)\alpha_1. \notag
    \end{align}
Notice that
\begin{align}
    \sqrt{1+\beta_1^2 \lambda \sigma_{U}^2} &= \left( 1 + \frac{\beta_1^{*2}\lambda \sigma_U^2}{\lambda^2 - \beta_1^{*2}\lambda \sigma_U^2} \right)^{1/2} \notag\\
    &= \left ( \frac{\lambda^2 - \beta_1^{*2}\lambda \sigma_U^2 + \beta_1^{*2}\lambda \sigma_U^2 }{\lambda^2 - \beta_1^{*2}\lambda \sigma_U^2}\right )^ {1/2} \notag \\
    &= \frac{\lambda}{[\lambda^2 - \beta_1^{*2}\lambda \sigma_U^2]^{1/2}} \label{eq7}.
\end{align}
It follows that \begin{align}
     \beta_0 &=\beta_0^* \sqrt{1+\beta_1^2 \lambda \sigma_{U}^2} - \beta_1(1-\lambda)\alpha_0 \notag\\
     &= \beta_0^* \frac{\lambda}{\sqrt{ \lambda^2 -\beta_1^{*2}\lambda \sigma_{U}^2 }} - \frac{\beta_1^* (1-\lambda)\alpha_0}{\sqrt{ \lambda^2 -\beta_1^{*2}\lambda \sigma_{U}^2 }} \notag \hspace{1cm} \text{(using equation~(\ref{eq7}))}\\ 
      &= \frac{\beta_0^* \lambda - \beta_1^* (1-\lambda)\alpha_0 }{\sqrt{ \lambda^2 -\beta_1^{*2}\lambda \sigma_{U}^2 }}, \label{beta0} \\
      \beta_2 &= \beta_2^* \sqrt{1+\beta_1^2 \lambda \sigma_{U}^2} - \beta_1(1-\lambda)\alpha_1 \notag \\
      &= \beta_2^* \frac{\lambda}{\sqrt{ \lambda^2 -\beta_1^{*2}\lambda \sigma_{U}^2 }} - \frac{\beta_1^* (1-\lambda)\alpha_1}{\sqrt{ \lambda^2 -\beta_1^{*2}\lambda \sigma_{U}^2 }}\notag \hspace{1cm} \text{(using equation~(\ref{eq7}))}\\
      &= \frac{\beta_2^* \lambda - \beta_1^* (1-\lambda)\alpha_1}{\sqrt{ \lambda^2 -\beta_1^{*2}\lambda \sigma_{U}^2 }}. \label{beta2}
\end{align}
Notice that \begin{align}
    \sqrt{1+\beta_1^2\sigma_M^2} &= \left( 1 + \frac{\beta_1^{*2}}{\lambda^2 -\beta_1^{*2}\lambda \sigma_{U}^2}\lambda\sigma_{M^*}^2\right)^{1/2} \notag \\
    &= \left(\frac{ \lambda^2 -\beta_1^{*2}\lambda \sigma_{U}^2 + \beta_1^{*2} \lambda\sigma_{M^*}^2}{\lambda^2 -\beta_1^{*2}\lambda \sigma_{U}^2} \right)^{1/2} \notag\\
    &= \left(\frac{ \lambda^2 -\beta_1^{*2}\lambda (1-\lambda)\sigma_{M^*}^2 + \beta_1^{*2} \lambda\sigma_{M^*}^2}{\lambda^2 -\beta_1^{*2}\lambda \sigma_{U}^2} \right)^{1/2} \notag \\
    &= \left(\frac{ \lambda^2 (1 + \beta_1^{*2} \sigma_{M^*}^2)}{\lambda^2 -\beta_1^{*2}\lambda \sigma_{U}^2} \right)^{1/2} \notag \\
    & = \frac{\lambda \sqrt{1 + \beta_1^{*2} \sigma_{M^*}^2}}{\sqrt{\lambda^2 -\beta_1^{*2}\lambda \sigma_{U}^2}} \label{eq8}.
\end{align}
To estimate the pure indirect effect or organic indirect effect relative to $a=0$, ${\mathbb{E}}[Y^{(0,I=1)}] - {\mathbb{E}}[Y^{(0)}] $, consider ${\mathbb{E}}[Y^{(0,I=1)}]$ where the shift to the left caused by the treatment is $\xi. $

\begin{scriptsize}

\begin{align} \label{A14}
  & {\mathbb{E}}[Y^{(0,I=1)}] \notag\\
  &= \sum_{c=0}^1 \left[ \Phi\left(\frac{{\beta}_0 + {\beta}_1({\alpha}_0 + {\alpha}_1c -\xi)+{\beta}_2c}{\sqrt{1+({\beta}_1{\sigma}_M)^2}} \right)  \right]P(C=c) \notag\\
    &= \sum_{c=0}^1 \Biggr[ \Phi\left( \frac{\sqrt{\lambda^2 -\beta_1^{*2}\lambda \sigma_{U}^2}}{\lambda \sqrt{1 + \beta_1^{*2} \sigma_{M^*}^2}} \left[ \frac{\beta_0^* \lambda - \beta_1^* (1-\lambda)\alpha_0 }{\sqrt{ \lambda^2 -\beta_1^{*2}\lambda \sigma_{U}^2 }} + \frac{\beta_1^* ({\alpha}_0 + {\alpha}_1c -\xi)}{\sqrt{ \lambda^2 -\beta_1^{*2}\lambda \sigma_{U}^2 }} + \frac{(\beta_2^* \lambda - \beta_1^* (1-\lambda)\alpha_1)c}{\sqrt{ \lambda^2 -\beta_1^{*2}\lambda \sigma_{U}^2 }}\right] \right) \Biggr]P(C=c) \notag\\
    &= \sum_{c=0}^1 \Biggr[ \Phi\left( \frac{1}{\lambda \sqrt{1 + \beta_1^{*2} \sigma_{M^*}^2}} \left[ \beta_0^* \lambda + \beta_1^* (-(1-\lambda)\alpha_0 +{\alpha}_0 + {\alpha}_1c -\xi  - (1-\lambda)\alpha_1c) + \beta_2^* \lambda c\right] \right) \Biggr]P(C=c) \notag \\
    &= \sum_{c=0}^1 \Biggr[ \Phi\left( \frac{1}{\lambda \sqrt{1 + \beta_1^{*2} \sigma_{M^*}^2}} \left[ \beta_0^* \lambda + \beta_1^* ( \lambda\alpha_0 + \lambda \alpha_1c -\xi) + \beta_2^* \lambda c\right] \right) \Biggr]P(C=c). \notag\\
\end{align}
\end{scriptsize}
where the second equality uses equations ~(\ref{beta1}), ~(\ref{beta0}), ~(\ref{beta2}) and ~(\ref{eq8}).

For ${\mathbb{E}}[Y^{(0)}]$, the same result holds after plugging in $\xi = 0$. \\
Starting from equation (\ref{A14}), the derivative with respect to $\alpha_0$ equals
\begin{equation}
\begin{split} 
    \frac{\partial \mathbb{E}[Y^{(0,I=1)}] }{\partial \alpha_0} &= \sum_{C=0}^{1} \Biggr[  \phi(.^1) \frac{\beta_1^* \lambda }{\lambda \sqrt{1 + \beta_1^{*2} \sigma_{M^*}^2}}\Biggr]P(C=c) \\
    &= \sum_{C=0}^{1} \Biggr[  \phi(.^1) \frac{\beta_1^*}{ \sqrt{1 + \beta_1^{*2} \sigma_{M^*}^2}}\Biggr] P(C=c), 
    \end{split}   
    \end{equation}     
    where  $$\phi(.^1) = \phi \left( \frac{\beta_0^* \lambda + \beta_1^* ( \lambda\alpha_0 + \lambda\alpha_1c - \xi) + \beta_2^* \lambda c}{\lambda \sqrt{1 + \beta_1^{*2} \sigma_{M^*}^2}} \right).$$

The derivative with respect to $\alpha_1$ equals
\begin{equation}
\begin{split} 
    \frac{\partial \mathbb{E}[Y^{(0,I=1)}]}{\partial \alpha_1} &= \sum_{C=0}^{1} \Biggr[  \phi(.^1) \frac{\beta_1^* \lambda c }{\lambda \sqrt{1 + \beta_1^{*2} \sigma_{M^*}^2}}\Biggr]P(C=c)\\
    &= \sum_{C=0}^{1} \Biggr[  \phi(.^1) \frac{\beta_1^* c}{ \sqrt{1 + \beta_1^{*2} \sigma_{M^*}^2}}\Biggr] P(C=c) .
\end{split}
\end{equation}

The derivative with respect to $\beta_0^*$ equals
\begin{equation}
\begin{split} 
    \frac{\partial \mathbb{E}[Y^{(0,I=1)}]}{\partial \beta_0^*} &= \sum_{C=0}^{1} \Biggr[  \phi(.^1) \frac{\lambda}{\lambda \sqrt{1 + \beta_1^{*2} \sigma_{M^*}^2}}\Biggr]P(C=c) \\
    &= \sum_{C=0}^{1} \Biggr[  \phi(.^1) \frac{1}{ \sqrt{1 + \beta_1^{*2} \sigma_{M^*}^2}}\Biggr] P(C=c). \\
\end{split}
\end{equation}

The derivative with respect to $\beta_1^*$ equals
\begin{footnotesize}
\begin{equation}
\begin{split} 
   & \frac{\partial \mathbb{E}[Y^{(0,I=1)}]}{\partial \beta_1^*} \\
   &= \sum_{C=0}^{1} \Biggr[ \phi(.^1) \lambda \sqrt{1 + \beta_1^{*2} \sigma_{M^*}^2} \Bigl(\lambda\alpha_0 + \lambda \alpha_1c -\xi\Bigl) - \Bigr[ \beta_0^* \lambda + \beta_1^* ( \lambda\alpha_0 + \lambda \alpha_1c -\xi) + \beta_2^* \lambda c\Bigr] \\ 
   & \hspace{10mm} \Bigr[\lambda(1 + \beta_1^{*2}\sigma_{M^*}^2)^{-\frac{1}{2}}\beta_1^*\sigma_{M^*}^2\Bigr]\Biggr] \cdot \frac{P(C=c)}{\lambda^2 (1 + \beta_1^{*2} \sigma_{M^*}^2)} \\
  &= \sum_{C=0}^{1} \Biggr[ \phi(.^1) \frac{(\lambda\alpha_0 + \lambda \alpha_1c -\xi) - \beta_1^*\sigma_{M^*}^2 \lambda \left( \beta_0^*  + \beta_2^* c\right)}{\lambda (1 + \beta_1^{*2} \sigma_{M^*}^2)^{3/2}}\Biggr]P(C=c).
\end{split}
\end{equation}
\end{footnotesize} 

The derivative with respect to $\beta_2^*$ equals
\begin{equation}
\begin{split} 
    \frac{\partial \mathbb{E}[Y^{(0,I=1)}]}{\partial \beta_2^*} &= \sum_{C=0}^{1} \Biggr[  \phi(.^1) \frac{\lambda c}{\lambda \sqrt{1 + \beta_1^{*2} \sigma_{M^*}^2}}\Biggr]P(C=c), \\
    &= \sum_{C=0}^{1} \Biggr[  \phi(.^1) \frac{c}{ \sqrt{1 + \beta_1^{*2} \sigma_{M^*}^2}}\Biggr] P(C=c). 
\end{split}
\end{equation}
For the derivative with respect to $\sigma_{M^*}^2$, note that $$\lambda = \frac{\sigma_M^2}{\sigma_{M^*}^2} = \frac{\sigma_{M^*}^2 - \sigma_U^2}{\sigma_{M^*}} = 1 - \frac{\sigma_U^2}{\sigma_{M^*}^2}.$$ Then from equation~(\ref{A14}), 
\begin{footnotesize}
\begin{equation}
\begin{split} 
    & {\mathbb{E}}[Y^{(0,I=1)}] \\
    &= \sum_{c=0}^1 \left[ \Phi\left( \frac{\left(1 + \beta_1^{*2} \sigma_{M^*}^2 \right)^{-\frac{1}{2}}}{\left ( 1 - \frac{\sigma_U^2}{\sigma_{M^*}^2}\right) } \left[ \left ( 1 - \frac{\sigma_U^2}{\sigma_{M^*}^2}\right) [\beta_0^* + \beta_1^* ( \alpha_0 + \alpha_1c) + \beta_2^* c ] - \beta_1^* \xi \right] \right) \right]P(C=c). \label{eq15}\\
\end{split}
\end{equation}
\end{footnotesize}
Using equation~(\ref{eq15}), the derivative with respect to $\sigma_{M^*}^2$ equals
\begin{scriptsize}
\begin{equation}
\begin{split} 
    & \frac{\mathbb{E}[Y^{(0,I=1)}]}{\partial \sigma_{M^*}^2} \\
    &= \sum_{C=0}^{1} \Biggl\{  \phi(.^1) \Biggr[\left ( 1 - \frac{\sigma_U^2}{\sigma_{M^*}^2}\right)\sqrt{1 + \beta_1^{*2} \sigma_{M^*}^2} [\beta_0^* + \beta_1^* ( \alpha_0 + \alpha_1c) + \beta_2^* c ]\left(\frac{\sigma_U^2}{\sigma_{M^*}^4} \right) \\
    & \hspace{5mm} - \Biggr[\left ( 1 - \frac{\sigma_U^2}{\sigma_{M^*}^2}\right) [\beta_0^* + \beta_1^* ( \alpha_0 + \alpha_1c) + \beta_2^* c ] - \beta_1^* \xi \Biggr] \\
     & \hspace{5mm} \Biggr[ \sqrt{1 + \beta_1^{*2} \sigma_{M^*}^2}\left( \frac{\sigma_U^2}{\sigma_{M^*}^4} \right) + \left ( 1 - \frac{\sigma_U^2}{\sigma_{M^*}^2}\right)\frac{1}{2}(1 + \beta_1^{*2}\sigma_{M^*}^2 )^{-1/2}\beta_1^{*2} \Biggr] \Biggr]\Biggl\} \frac{P(C=c)}{\left ( 1 - \frac{\sigma_U^2}{\sigma_{M^*}^2}\right)^2 (1 + \beta_1^{*2}\sigma_{M^*}^2)} \\
     &= \sum_{C=0}^{1} \Biggl\{   \phi(.^1) \Biggr[ \lambda \sqrt{1 + \beta_1^{*2} \sigma_{M^*}^2} [\beta_0^* + \beta_1^* ( \alpha_0 + \alpha_1c) + \beta_2^* c ]\left(\frac{\sigma_U^2}{\sigma_{M^*}^4} \right) \\
     & \hspace{5mm} - \Biggl(\lambda [\beta_0^* + \beta_1^* ( \alpha_0 + \alpha_1c) + \beta_2^* c ] - \beta_1^* \xi \Biggl)
      \\
     & \hspace{5mm}  \Biggl( \sqrt{1 + \beta_1^{*2} \sigma_{M^*}^2}\left( \frac{\sigma_U^2}{\sigma_{M^*}^4} \right) + \frac{\lambda \beta_1^{*2}}{2\sqrt{1 + \beta_1^{*2}\sigma_{M^*}^2}} \Biggl)\Biggr]\Biggl\}\frac{P(C=c)}{\lambda^2 (1 + \beta_1^{*2}\sigma_{M^*}^2)} .
\end{split}
\end{equation}
\end{scriptsize}

\newpage

\section{RESULTS} \label{Appendix_B} 

Appendix~\ref{Appendix_B} provides additional results.
\FloatBarrier

\begin{table}[H]
    \centering
        \caption{Estimates and 95\% confidence intervals for $\hat{\alpha}_0, \hat{\alpha}_1, \hat{\sigma}_M^2, \hat{\beta}_0, \hat{\beta}_1, \hat{\beta}_2$}
    \begin{tabular}{c c c c}
    \hline
        \makecell{HIV persistence \\ 
     measure} & Coefficient & Point estimate & \makecell{95\% CI} \\ \hline
        caRNA & $\hat{\alpha}_0$ & $1.37$ & $(1.04, 1.71)$ \\ 
        & $\hat{\alpha}_1$ & $0.95$ & $(0.57, 1.34)$\\ 
         & $\hat{\sigma}_M^2 $ & $0.62$ & $(0.37, 0.87)$ \\ 
          & $\hat{\beta}_0 $ & $0.60$ & $(-0.07, 1.27)$ \\ \
            & $\hat{\beta}_1 $ & $-0.70$ & $(-1.06, -0.35)$ \\ 
              & $\hat{\beta}_2 $ & $0.55$ & $(-0.11, 1.20) $ \\ \hline
        SCA & $\hat{\alpha}_0$ & $-0.19$ & $(-0.63, 0.25)$ \\ 
        & $\hat{\alpha}_1$ & $0.12$ & $(-0.37, 0.62)$ \\ 
         & $\hat{\sigma}_M^2 $ & $0.70$ & $(0.27, 1.13)$ \\ 
          & $\hat{\beta}_0 $ & $-0.27$ & $(-0.79, 0.25)$ \\ \
            & $\hat{\beta}_1 $ & $-0.28$ & $(-0.59, 0.03)$ \\ 
              & $\hat{\beta}_2 $ & $-0.15$ & $(-0.76, 0.45)$ \\ \hline
    \end{tabular}
    \label{tab:fitted_models}
\end{table}

\newpage
\begin{table*}[h]
\caption{Sensitivity analysis: Pure indirect effects/organic indirect effects relative to no treatment of curative HIV treatments that shift the distribution of HIV persistence measures on the $log_{10}$ scale downwards - different assay limits}
\label{App_ie_results}
\begin{threeparttable}
\begin{tabular}{@{}p{2.4cm} >{\centering\arraybackslash}p{2cm} c >{\centering\arraybackslash}p{2.1cm} >{\centering\arraybackslash}p{2.1cm} >{\centering\arraybackslash}p{2.7cm}@{}}
\toprule
\makecell{HIV Persistence \\ Measure} & \makecell{Mediator \\ Shift\tnote{a}} & Week\tnote{b} & 
\multicolumn{2}{c}{Indirect Effect\tnote{c}} & \makecell{95\% CI\tnote{d}} \\
\cmidrule(lr){4-5}
& & & measurement error ignored & measurement error adjusted & \\
\midrule
     caRNA\tnote{e,f} & 0.5 & 4   & 14.5\% & 18.2\% & (8.7\%, 27.6\%)\\
     caRNA            & 1.0 & 4   & 29.1\% & 35.8\% & (19.4\%, 52.4\%)\\
     caRNA            & 1.5 & 4   & 42.0\% & 49.6\% & (32.1\%, 67.0\%)\\
     caRNA            & 2.0 & 4   & 51.8\% & 58.3\% & (44.1\%, 72.5\%)\\ \hline
     SCA\tnote{g,h} & 0.5 & 4  & 4.1\% & 6.4\%  & (-4.6\%, 17.3\%)\\
     SCA            & 1.0 & 4  & 8.4\% & 12.9\% & (-9.4\%, 35.2\%)\\
     SCA            & 1.5 & 4  & 12.7\% & 19.5\% & (-13.5\%, 52.6\%)\\
     SCA            & 2.0 & 4  & 17.0\% & 25.9\% & (-16.5\%, 68.3\%)\\ \hline
     caRNA\tnote{e,f} & 0.5 & 8  & 10.8\% & 14.2\% & (3.2\%, 25.2\%)\\
     caRNA            & 1.0 & 8  & 25.6\% & 33.9\% & (8.3\%, 59.5\%)\\
     caRNA            & 1.5 & 8  & 42.4\% & 54.6\% & (20.3\%, 89.0\%)\\
     caRNA            & 2.0 & 8  & 58.4\% & 71.1\% & (40.1\%, 100.0\%)\\ \hline
     SCA\tnote{g,h}  & 0.5 & 8   & 7.3\% & 12.1\%  & (0.5\%, 23.7\%)\\
     SCA             & 1.0 & 8   & 17.1\% & 29.4\%  & (0.7\%, 58.0\%)\\
     SCA             & 1.5 & 8   & 28.8\% & 48.8\%  & (6.4\%, 89.3\%)\\
     SCA             & 2.0 & 8   & 41.6\% & 66.1\%  & (22.2\%, 100.0\%) \\ \hline
    \end{tabular}
        \begin{footnotesize}
    The estimated probability of virologic suppression through ATI week 4 without curative treatment was 36/104 (or 34.6\%) and 13/104 (or 12.5\%) through ATI week 8 in participants with caRNA measured. \\
     The estimated probability of virologic suppression through ATI week 4 without curative treatment was 32/88 (or 36.4\%) and 10/88 (or 11.4\%) through ATI week 8 in participants with SCA measured. 
    
    \tnote{a} Curative treatment-induced downward shift of the viral persistence measure distribution, given the common cause, on the $log_{10}$ scale (compared to no curative treatment).
     
     \tnote{b} Viral rebound in the first 4 weeks and first 8 weeks after ART interruption. 
     
    \tnote{c} Pure or organic indirect effect relative to no treatment: the difference in probability of virologic suppression that is mediated through the HIV persistence measure.
    
    \tnote{d} 95\% Confidence Interval (95\% CI) calculated using the Delta method.
    
    \tnote{e} Cell-associated HIV RNA, on ART. All 104 participants had caRNA measured. Assay limit across all studies: 92 copies per million CD4+ T cells.
    
    \tnote{f} $log_{10}$ caRNA: estimated measurement error SD = 0.29 (\cite{Scully2022}).
    
    \tnote{g} Single-copy plasma HIV RNA, on ART. Restricted to the 88 participants with SCA measured. Assay limit across all studies: 1 copy per ml.
    
    \tnote{h} $log_{10}$ SCA: estimated measurement error SD = 0.47 (\cite{Riddler2016}).
    \end{footnotesize}
\end{threeparttable}

\end{table*}

\begin{table}[H]
\caption{Sensitivity analysis: pure indirect effects/organic indirect effects relative to no treatment of curative HIV treatments that shift the distribution of HIV persistence measures downwards using logit link - measurement error adjusted}
\label{ie_results_logit}
\begin{threeparttable}
    \begin{tabular}{@{}l >{\centering\arraybackslash}p{2cm} c c >{\centering\arraybackslash}p{3cm} c c c@{}}
    \hline
     \makecell{HIV persistence \\ 
     measure} & \makecell{Mediator \\ Shift\tnote{a}} & Week\tnote{b} &  \makecell{Indirect effect\tnote{c} } & \makecell{95\% CI\tnote{d}}
 \\ \hline
     caRNA\tnote{e,f} & 0.5 & 4   & 11.6\% & (5.5\%, 17.4\%)\\
     caRNA            & 1.0 & 4   & 23.2\% & (11.6\%, 34.8\%)\\
     caRNA            & 1.5 & 4   & 34.2\% & (18.5\%, 49.9\%)\\
     caRNA            & 2.0 & 4   & 43.7\% & (26.4\%, 61.1\%)\\ \hline
     SCA \tnote{g,h} & 0.5 & 4  & 5.1\%  & (-2.2\%, 12.4\%)\\
     SCA            & 1.0 & 4  & 10.4\% & (-4.6\%, 25.3\%)\\
     SCA            & 1.5 & 4  & 15.7\% & (-6.7\%, 38.0\%)\\
     SCA            & 2.0 & 4  & 20.9\% & (-8.4\%, 50.2\%)\\ \hline
     caRNA\tnote{e,f} & 0.5 & 8   & 8.0\% & (2.4\%, 13.6\%)\\
     caRNA            & 1.0 & 8   & 18.6\% & (4.9\%, 32.3\%)\\
     caRNA            & 1.5 & 8   & 31.1\% & (8.5\%, 53.7\%)\\
     caRNA            & 2.0 & 8   & 44.5\% & (14.5\%, 74.4\%)\\ \hline
     SCA\tnote{g,h}  & 0.5 & 8 & 9.2\%  & (1.5\%, 16.9\%)\\
     SCA             & 1.0 & 8 & 22.1\%  & (2.8\%, 41.3\%)\\
     SCA             & 1.5 & 8 & 37.5\%  & (6.0\%, 68.9\%)\\
     SCA             & 2.0 & 8 & 53.2\%  & (14.0\%, 92.4\%) \\ \hline
    \end{tabular}
       \begin{footnotesize}
    The estimated probability of virologic suppression through ATI week 4 without curative treatment was 36/104 (or 34.6\%) and 13/104 (or 12.5\%) through ATI week 8 in 104 participants with caRNA measured. \\
     The estimated probability of virologic suppression through ATI week 4 without curative treatment was 32/88 (or 36.4\%) and 10/88 (or 11.4\%) through ATI week 8 in 88 participants with SCA measured. 
    
    \tnote{a} Curative treatment-induced downward shift of the viral persistence measure distribution, given the common cause, on the $log_{10}$ scale (compared to no curative treatment).
     
     \tnote{b} Viral rebound in the first 4 weeks and first 8 weeks after ART interruption. 
     
    \tnote{c} Pure or organic indirect effect relative to no treatment: the difference in probability of virologic suppression that is mediated through the HIV persistence measure.
    
    \tnote{d} 95\% Confidence Interval (95\% CI) calculated using the Delta method.
    
    \tnote{e} Cell-associated HIV RNA, on ART. All 104 participants had caRNA measured. Assay limit of ACTG A5345: 28 copies per million CD4+ T cells. Assay limit of remaining studies: 92 copies per million CD4+ T cells.
    
    \tnote{f} $log_{10}$ caRNA: estimated measurement error SD = 0.29 (\cite{Scully2022}).
    
    \tnote{g} Single-copy plasma HIV RNA, on ART. Restricted to the 88 participants with SCA measured. Assay limit of ACTG A5345: 0.56 copies per ml. Assay limit of remaining studies: 1 copy per ml.
    
    \tnote{h} $log_{10}$ SCA: estimated measurement error SD = 0.47 (\cite{Riddler2016}).
    \end{footnotesize}
\end{threeparttable}
\end{table}

\begin{table}[H]
\caption{Bootstrap confidence intervals: pure indirect effects/organic indirect effects relative to no treatment of curative HIV treatments that shift the distribution of HIV persistence measures on the $log_{10}$ scale downwards.}
\label{ie_results_bootstrap}
\begin{threeparttable}
    \begin{tabular}{@{}l >{\centering\arraybackslash}p{2cm} c c >{\centering\arraybackslash}p{3cm} c c c@{}}
    \hline
     \makecell{HIV persistence \\ 
     measure} & \makecell{Mediator \\ Shift\tnote{a}} & Week\tnote{b} &  \makecell{Indirect effect\tnote{c} } & \makecell{95\% CI\tnote{d}}
 \\ \hline
     CA HIV RNA\tnote{e,f} & 0.5 & 4   & 11.9\% & (5.9\%, 17.6\%)\\
     CA HIV RNA            & 1.0 & 4   & 24.1\% & (15.6\%, 37.7\%)\\
     CA HIV RNA            & 1.5 & 4   & 35.5\% & (23.4\%, 50.2\%)\\
     CA HIV RNA            & 2.0 & 4   & 45.2\% & (30.9\%, 58.2\%)\\ \hline
     SCA HIV RNA\tnote{g,h} & 0.5 & 4  & 5.3\%  & (-2.4\%, 13.0\%)\\
     SCA HIV RNA            & 1.0 & 4  & 10.7\% & (-4.8\%, 25.8\%)\\
     SCA HIV RNA            & 1.5 & 4  & 16.2\% & (-7.2\%, 37.0\%)\\
     SCA HIV RNA            & 2.0 & 4  & 21.6\% & (-9.5\%, 46.3\%)\\ \hline
    \end{tabular}
       \begin{footnotesize}
    The estimated probability of virologic suppression through week 4 without curative treatment was 36/104 (or 34.6\%) and 13/104 (or 12.5\%) at week 8 for CA HIV RNA. 

    \tnote{a} Curative treatment-induced a downward shift of the viral persistence measure distribution, given the common cause, on the $log_{10}$ scale (compared to no curative treatment).
     
     \tnote{b} Analyzing viral rebound in the first 4 weeks and first 8 weeks after ART interruption. 
     
    \tnote{c} Pure (or organic indirect effect relative to no treatment): the difference in probability of virologic suppression that is mediated. 
    
    \tnote{d} 95\% Confidence Interval (95\% CI) calculated by bootstrap with 2000 replicates using Efron's percentile method.
    
    \tnote{e} Cell-associated HIV RNA, on ART. All 104 participants had caRNA measured. Assay limit of ACTG A5345: 28 copies per million CD4+ T cells. Assay limit of remaining studies: 92 copies per million CD4+ T cells.
    
    \tnote{f} $log_{10}$ caRNA: estimated measurement error SD = 0.29 (\cite{Scully2022}).
    
    \tnote{g} Single-copy plasma HIV RNA, on ART. Restricted to the 88 participants with SCA measured. Assay limit of ACTG A5345: 0.56 copies per ml. Assay limit of remaining studies: 1 copy per ml.
    
    \tnote{h} $log_{10}$ SCA: estimated measurement error SD = 0.47 (\cite{Riddler2016}).
    \end{footnotesize}
\end{threeparttable}
\end{table}

%% file: Manuscript.bbl
\begin{thebibliography}{27}

\bibitem[\protect\citeauthoryear{Baron and Kenny}{1986}]{Baron1986}
\begin{barticle}[author]
\bauthor{\bsnm{Baron},~\bfnm{Reuben~M.}\binits{R.~M.}} \AND \bauthor{\bsnm{Kenny},~\bfnm{David~A.}\binits{D.~A.}}
(\byear{1986}).
\btitle{The Moderator-Mediator Variable Distinction in Social Psychological Research. Conceptual, Strategic, and Statistical Considerations}.
\bjournal{Journal of Personality and Social Psychology}
\bvolume{51}.
\bdoi{10.1037/0022-3514.51.6.1173}
\end{barticle}
\endbibitem

\bibitem[\protect\citeauthoryear{Chernofsky, Bosch and Lok}{2024}]{Chernofsky}
\begin{barticle}[author]
\bauthor{\bsnm{Chernofsky},~\bfnm{Ariel}\binits{A.}}, \bauthor{\bsnm{Bosch},~\bfnm{Ronald~J.}\binits{R.~J.}} \AND \bauthor{\bsnm{Lok},~\bfnm{Judith~J.}\binits{J.~J.}}
(\byear{2024}).
\btitle{Causal mediation analysis with mediator values below an assay limit}.
\bjournal{Statistics in Medicine}
\bvolume{43}
\bpages{2299-2313}.
\bdoi{https://doi.org/10.1002/sim.10065}
\end{barticle}
\endbibitem

\bibitem[\protect\citeauthoryear{Deeks et~al.}{2009}]{Deeks}
\begin{barticle}[author]
\bauthor{\bsnm{Deeks},~\bfnm{Steven~G.}\binits{S.~G.}}, \bauthor{\bsnm{Gange},~\bfnm{Stephen~J.}\binits{S.~J.}}, \bauthor{\bsnm{Kitahata},~\bfnm{Mari~M.}\binits{M.~M.}}, \bauthor{\bsnm{Saag},~\bfnm{Michael~S.}\binits{M.~S.}}, \bauthor{\bsnm{Justice},~\bfnm{Amy~C.}\binits{A.~C.}}, \bauthor{\bsnm{Hogg},~\bfnm{Robert~S.}\binits{R.~S.}}, \bauthor{\bsnm{Eron},~\bfnm{Joseph~J.}\binits{J.~J.}}, \bauthor{\bsnm{Brooks},~\bfnm{John~T.}\binits{J.~T.}}, \bauthor{\bsnm{Rourke},~\bfnm{Sean~B.}\binits{S.~B.}}, \bauthor{\bsnm{Gill},~\bfnm{M.~John}\binits{M.~J.}}, \bauthor{\bsnm{Bosch},~\bfnm{Ronald~J.}\binits{R.~J.}}, \bauthor{\bsnm{Benson},~\bfnm{Constance~A.}\binits{C.~A.}}, \bauthor{\bsnm{Collier},~\bfnm{Ann~C.}\binits{A.~C.}}, \bauthor{\bsnm{Martin},~\bfnm{Jeffrey~N.}\binits{J.~N.}}, \bauthor{\bsnm{Klein},~\bfnm{Marina~B.}\binits{M.~B.}}, \bauthor{\bsnm{Jacobson},~\bfnm{Lisa~P.}\binits{L.~P.}}, \bauthor{\bsnm{Rodriguez},~\bfnm{Benigno}\binits{B.}}, \bauthor{\bsnm{Sterling},~\bfnm{Timothy~R.}\binits{T.~R.}},
  \bauthor{\bsnm{Kirk},~\bfnm{Gregory~D.}\binits{G.~D.}}, \bauthor{\bsnm{Napravnik},~\bfnm{Sonia}\binits{S.}}, \bauthor{\bsnm{Rachlis},~\bfnm{Anita~R.}\binits{A.~R.}}, \bauthor{\bsnm{Calzavara},~\bfnm{Liviana~M.}\binits{L.~M.}}, \bauthor{\bsnm{Horberg},~\bfnm{Michael~A.}\binits{M.~A.}}, \bauthor{\bsnm{Silverberg},~\bfnm{Michael~J.}\binits{M.~J.}}, \bauthor{\bsnm{Gebo},~\bfnm{Kelly~A.}\binits{K.~A.}}, \bauthor{\bsnm{Kushel},~\bfnm{Margot~B.}\binits{M.~B.}}, \bauthor{\bsnm{Goedert},~\bfnm{James~J.}\binits{J.~J.}}, \bauthor{\bsnm{McKaig},~\bfnm{Rosemary~G.}\binits{R.~G.}}, \bauthor{\bsnm{Moore},~\bfnm{Richard~D.}\binits{R.~D.}}, \bauthor{\bparticle{on} \bsnm{Research},~\bfnm{North American AIDS Cohort~Collaboration}\binits{N.~A. A. C.~C.}} \AND \bauthor{\bsnm{Design}}
(\byear{2009}).
\btitle{Trends in Multidrug Treatment Failure and Subsequent Mortality among Antiretroviral Therapy-Experienced Patients with HIV Infection in North America}.
\bjournal{Clinical Infectious Diseases}
\bvolume{49}
\bpages{1582-1590}.
\bdoi{10.1086/644768}
\end{barticle}
\endbibitem

\bibitem[\protect\citeauthoryear{Gunst et~al.}{2025}]{Gunst2025}
\begin{barticle}[author]
\bauthor{\bsnm{Gunst},~\bfnm{Jesper~D.}\binits{J.~D.}}, \bauthor{\bsnm{Gohil},~\bfnm{Jesal}\binits{J.}}, \bauthor{\bsnm{Li},~\bfnm{Johanthan~Z.}\binits{J.~Z.}}, \bauthor{\bsnm{Bosch},~\bfnm{Ronald~J.}\binits{R.~J.}}, \bauthor{\bsnm{Catherine~Seamon},~\bfnm{Andrea~White}\binits{A.~W.}}, \bauthor{\bsnm{Chun},~\bfnm{Tae-Wook}\binits{T.-W.}}, \bauthor{\bsnm{Mothe},~\bfnm{Beatriz}\binits{B.}}, \bauthor{\bsnm{Gittens},~\bfnm{Kathleen}\binits{K.}}, \bauthor{\bsnm{Praiss},~\bfnm{Lauren}\binits{L.}}, \bauthor{\bsnm{Scheerder},~\bfnm{Marie-Angélique~De}\binits{M.-A.~D.}}, \bauthor{\bsnm{Vandekerckhove},~\bfnm{Linos}\binits{L.}}, \bauthor{\bsnm{Escandón},~\bfnm{Kevin}\binits{K.}}, \bauthor{\bsnm{Thorkelson},~\bfnm{Ann}\binits{A.}}, \bauthor{\bsnm{Schacker},~\bfnm{Timothy}\binits{T.}}, \bauthor{\bsnm{SenGupta},~\bfnm{Devi}\binits{D.}}, \bauthor{\bsnm{Brander},~\bfnm{Christian}\binits{C.}}, \bauthor{\bsnm{Papasavvas},~\bfnm{Emmanouil}\binits{E.}}, \bauthor{\bsnm{Montaner},~\bfnm{Luis~J.}\binits{L.~J.}},
  \bauthor{\bsnm{Martinez-Picado},~\bfnm{Javier}\binits{J.}}, \bauthor{\bsnm{Calin},~\bfnm{Ruxandra}\binits{R.}}, \bauthor{\bsnm{Castagna},~\bfnm{Antonella}\binits{A.}}, \bauthor{\bsnm{Muccini},~\bfnm{Camilla}\binits{C.}}, \bauthor{\bparticle{de} \bsnm{Jong},~\bfnm{Wesley}\binits{W.}}, \bauthor{\bsnm{Leal},~\bfnm{Lorna}\binits{L.}}, \bauthor{\bsnm{Garcia},~\bfnm{Felipe}\binits{F.}}, \bauthor{\bsnm{Gruters},~\bfnm{Rob~A.}\binits{R.~A.}}, \bauthor{\bsnm{Tipoe},~\bfnm{Timothy}\binits{T.}}, \bauthor{\bsnm{Frater},~\bfnm{John}\binits{J.}}, \bauthor{\bsnm{Søgaard},~\bfnm{Ole~S.}\binits{O.~S.}} \AND \bauthor{\bsnm{Fidler},~\bfnm{Sarah}\binits{S.}}
(\byear{2025}).
\btitle{Time to HIV viral rebound and frequency of post-treatment control after analytical interruption of antiretroviral therapy: an individual data-based meta-analysis of 24 prospective studies}.
\bjournal{Nature Communications}
\bvolume{16}
\bpages{906}.
\bdoi{10.1038/s41467-025-56116-1}
\end{barticle}
\endbibitem

\bibitem[\protect\citeauthoryear{Günthard et~al.}{2018}]{10.1093/cid/ciy463}
\begin{barticle}[author]
\bauthor{\bsnm{Günthard},~\bfnm{Huldrych~F}\binits{H.~F.}}, \bauthor{\bsnm{Calvez},~\bfnm{Vincent}\binits{V.}}, \bauthor{\bsnm{Paredes},~\bfnm{Roger}\binits{R.}}, \bauthor{\bsnm{Pillay},~\bfnm{Deenan}\binits{D.}}, \bauthor{\bsnm{Shafer},~\bfnm{Robert~W}\binits{R.~W.}}, \bauthor{\bsnm{Wensing},~\bfnm{Annemarie~M}\binits{A.~M.}}, \bauthor{\bsnm{Jacobsen},~\bfnm{Donna~M}\binits{D.~M.}} \AND \bauthor{\bsnm{Richman},~\bfnm{Douglas~D}\binits{D.~D.}}
(\byear{2018}).
\btitle{Human Immunodeficiency Virus Drug Resistance: 2018 Recommendations of the International Antiviral Society–USA Panel}.
\bjournal{Clinical Infectious Diseases}
\bvolume{68}
\bpages{177-187}.
\bdoi{10.1093/cid/ciy463}
\end{barticle}
\endbibitem

\bibitem[\protect\citeauthoryear{Hardnett et~al.}{2009}]{Hardnett2009}
\begin{barticle}[author]
\bauthor{\bsnm{Hardnett},~\bfnm{Felicia~P.}\binits{F.~P.}}, \bauthor{\bsnm{Pals},~\bfnm{Sherri~L.}\binits{S.~L.}}, \bauthor{\bsnm{Borkowf},~\bfnm{Craig~B.}\binits{C.~B.}}, \bauthor{\bsnm{Parsons},~\bfnm{Jeffrey}\binits{J.}}, \bauthor{\bsnm{Gomez},~\bfnm{Cynthia}\binits{C.}} \AND \bauthor{\bsnm{O'Leary},~\bfnm{Ann}\binits{A.}}
(\byear{2009}).
\btitle{Assessing mediation in HIV intervention studies}.
\bjournal{Public Health Reports}
\bvolume{124}.
\bdoi{10.1177/003335490912400217}
\end{barticle}
\endbibitem

\bibitem[\protect\citeauthoryear{Herath, Bosch and Lok}{2025}]{Herath2025}
\begin{barticle}[author]
\bauthor{\bsnm{Herath},~\bfnm{Vindyani}\binits{V.}}, \bauthor{\bsnm{Bosch},~\bfnm{Ronald~J}\binits{R.~J.}} \AND \bauthor{\bsnm{Lok},~\bfnm{Judith}\binits{J.}}
(\byear{2025}).
\btitle{Supplement to "Indirect effect on viral suppression of an HIV curative treatment that reduces viral persistence: mediators subject to an assay limit and measurement error"}.
\end{barticle}
\endbibitem

\bibitem[\protect\citeauthoryear{Hill et~al.}{2014}]{Hill2014}
\begin{barticle}[author]
\bauthor{\bsnm{Hill},~\bfnm{Alison~L.}\binits{A.~L.}}, \bauthor{\bsnm{Rosenbloom},~\bfnm{Daniel I.~S.}\binits{D.~I.~S.}}, \bauthor{\bsnm{Fu},~\bfnm{Feng}\binits{F.}}, \bauthor{\bsnm{Nowak},~\bfnm{Martin~A.}\binits{M.~A.}} \AND \bauthor{\bsnm{Siliciano},~\bfnm{Robert~F.}\binits{R.~F.}}
(\byear{2014}).
\btitle{Predicting the outcomes of treatment to eradicate the latent reservoir for HIV-1}.
\bjournal{Proceedings of the National Academy of Sciences of the United States of America}
\bvolume{111}.
\bdoi{10.1073/pnas.1406663111}
\end{barticle}
\endbibitem

\bibitem[\protect\citeauthoryear{Imai, Keele and Yamamoto}{2010}]{Imai2010}
\begin{barticle}[author]
\bauthor{\bsnm{Imai},~\bfnm{Kosuke}\binits{K.}}, \bauthor{\bsnm{Keele},~\bfnm{Luke}\binits{L.}} \AND \bauthor{\bsnm{Yamamoto},~\bfnm{Teppei}\binits{T.}}
(\byear{2010}).
\btitle{Identification, inference and sensitivity analysis for causal mediation effects}.
\bjournal{Statistical Science}
\bvolume{25}.
\bdoi{10.1214/10-STS321}
\end{barticle}
\endbibitem

\bibitem[\protect\citeauthoryear{Li et~al.}{2016}]{Li2016}
\begin{barticle}[author]
\bauthor{\bsnm{Li},~\bfnm{Jonathan~Z.}\binits{J.~Z.}}, \bauthor{\bsnm{Etemad},~\bfnm{Behzad}\binits{B.}}, \bauthor{\bsnm{Ahmed},~\bfnm{Hayat}\binits{H.}}, \bauthor{\bsnm{Aga},~\bfnm{Evgenia}\binits{E.}}, \bauthor{\bsnm{Bosch},~\bfnm{Ronald~J.}\binits{R.~J.}}, \bauthor{\bsnm{Mellors},~\bfnm{John~W.}\binits{J.~W.}}, \bauthor{\bsnm{Kuritzkes},~\bfnm{Daniel~R.}\binits{D.~R.}}, \bauthor{\bsnm{Lederman},~\bfnm{Michael~M.}\binits{M.~M.}}, \bauthor{\bsnm{Para},~\bfnm{Michael}\binits{M.}} \AND \bauthor{\bsnm{Gandhi},~\bfnm{Rajesh~T.}\binits{R.~T.}}
(\byear{2016}).
\btitle{The size of the expressed HIV reservoir predicts timing of viral rebound after treatment interruption}.
\bjournal{AIDS}
\bvolume{30}.
\bdoi{10.1097/QAD.0000000000000953}
\end{barticle}
\endbibitem

\bibitem[\protect\citeauthoryear{Li et~al.}{2022}]{Li2022}
\begin{barticle}[author]
\bauthor{\bsnm{Li},~\bfnm{Jonathan~Z.}\binits{J.~Z.}}, \bauthor{\bsnm{Aga},~\bfnm{Evgenia}\binits{E.}}, \bauthor{\bsnm{Bosch},~\bfnm{Ronald~J.}\binits{R.~J.}}, \bauthor{\bsnm{Pilkinton},~\bfnm{Mark}\binits{M.}}, \bauthor{\bsnm{Kroon},~\bfnm{Eugène}\binits{E.}}, \bauthor{\bsnm{Maclaren},~\bfnm{Lynsay}\binits{L.}}, \bauthor{\bsnm{Keefer},~\bfnm{Michael}\binits{M.}}, \bauthor{\bsnm{Fox},~\bfnm{Lawrence}\binits{L.}}, \bauthor{\bsnm{Barr},~\bfnm{Liz}\binits{L.}}, \bauthor{\bsnm{Acosta},~\bfnm{Edward}\binits{E.}}, \bauthor{\bsnm{Ananworanich},~\bfnm{Jintanat}\binits{J.}}, \bauthor{\bsnm{Coombs},~\bfnm{Robert}\binits{R.}}, \bauthor{\bsnm{Mellors},~\bfnm{John~W.}\binits{J.~W.}}, \bauthor{\bsnm{Landay},~\bfnm{Alan~L.}\binits{A.~L.}}, \bauthor{\bsnm{Macatangay},~\bfnm{Bernard}\binits{B.}}, \bauthor{\bsnm{Deeks},~\bfnm{Steven}\binits{S.}}, \bauthor{\bsnm{Gandhi},~\bfnm{Rajesh~T.}\binits{R.~T.}} \AND \bauthor{\bsnm{Smith},~\bfnm{Davey~M.}\binits{D.~M.}}
(\byear{2022}).
\btitle{Time to Viral Rebound After Interruption of Modern Antiretroviral Therapies}.
\bjournal{Clinical Infectious Diseases}
\bvolume{74}.
\bdoi{10.1093/cid/ciab541}
\end{barticle}
\endbibitem

\bibitem[\protect\citeauthoryear{Li et~al.}{2024}]{Lietal_2024}
\begin{barticle}[author]
\bauthor{\bsnm{Li},~\bfnm{Jonathan~Z.}\binits{J.~Z.}}, \bauthor{\bsnm{Melberg},~\bfnm{Meghan}\binits{M.}}, \bauthor{\bsnm{Kittilson},~\bfnm{Autumn}\binits{A.}}, \bauthor{\bsnm{Abdel-Mohsen},~\bfnm{Mohamed}\binits{M.}}, \bauthor{\bsnm{Li},~\bfnm{Yijia}\binits{Y.}}, \bauthor{\bsnm{Aga},~\bfnm{Evgenia}\binits{E.}}, \bauthor{\bsnm{Bosch},~\bfnm{Ronald~J.}\binits{R.~J.}}, \bauthor{\bsnm{Wonderlich},~\bfnm{Elizabeth~R.}\binits{E.~R.}}, \bauthor{\bsnm{Kinslow},~\bfnm{Jennifer}\binits{J.}}, \bauthor{\bsnm{Giron},~\bfnm{Leila~B.}\binits{L.~B.}}, \bauthor{\bsnm{Germanio},~\bfnm{Clara~Di}\binits{C.~D.}}, \bauthor{\bsnm{Pilkinton},~\bfnm{Mark}\binits{M.}}, \bauthor{\bsnm{MacLaren},~\bfnm{Lynsay}\binits{L.}}, \bauthor{\bsnm{Keefer},~\bfnm{Michael}\binits{M.}}, \bauthor{\bsnm{Fox},~\bfnm{Lawrence}\binits{L.}}, \bauthor{\bsnm{Barr},~\bfnm{Liz}\binits{L.}}, \bauthor{\bsnm{Acosta},~\bfnm{Edward}\binits{E.}}, \bauthor{\bsnm{Ananworanich},~\bfnm{Jintanat}\binits{J.}}, \bauthor{\bsnm{Coombs},~\bfnm{Robert}\binits{R.}},
  \bauthor{\bsnm{Mellors},~\bfnm{John}\binits{J.}}, \bauthor{\bsnm{Deeks},~\bfnm{Steven}\binits{S.}}, \bauthor{\bsnm{Gandhi},~\bfnm{Rajesh~T.}\binits{R.~T.}}, \bauthor{\bsnm{Busch},~\bfnm{Michael}\binits{M.}}, \bauthor{\bsnm{Landay},~\bfnm{Alan}\binits{A.}}, \bauthor{\bsnm{Macatangay},~\bfnm{Bernard}\binits{B.}}, \bauthor{\bsnm{Smith},~\bfnm{Davey~M.}\binits{D.~M.}} \AND \bauthor{\bparticle{for~the} \bsnm{AIDS Clinical Trials Group A5345 Study~Team}}
(\byear{2024}).
\btitle{Predictors of HIV rebound differ by timing of antiretroviral therapy initiation}.
\bjournal{JCI Insight}
\bvolume{9}.
\bdoi{10.1172/jci.insight.173864}
\end{barticle}
\endbibitem

\bibitem[\protect\citeauthoryear{Lok}{2016}]{Lok2016}
\begin{barticle}[author]
\bauthor{\bsnm{Lok},~\bfnm{Judith~J.}\binits{J.~J.}}
(\byear{2016}).
\btitle{Defining and estimating causal direct and indirect effects when setting the mediator to specific values is not feasible}.
\bjournal{Statistics in Medicine}
\bvolume{35}.
\bdoi{10.1002/sim.6990}
\end{barticle}
\endbibitem

\bibitem[\protect\citeauthoryear{Lok and Bosch}{2021}]{Lok2021}
\begin{barticle}[author]
\bauthor{\bsnm{Lok},~\bfnm{Judith~J.}\binits{J.~J.}} \AND \bauthor{\bsnm{Bosch},~\bfnm{Ronald~J.}\binits{R.~J.}}
(\byear{2021}).
\btitle{Causal Organic Indirect and Direct Effects: Closer to the Original Approach to Mediation Analysis, with a Product Method for Binary Mediators}.
\bjournal{Epidemiology}
\bvolume{32}
\bpages{412-420}.
\bdoi{10.1097/EDE.0000000000001339}
\end{barticle}
\endbibitem

\bibitem[\protect\citeauthoryear{Matza et~al.}{2017}]{matza2017risks}
\begin{barticle}[author]
\bauthor{\bsnm{Matza},~\bfnm{Louis~S}\binits{L.~S.}}, \bauthor{\bsnm{Chung},~\bfnm{Karen~C}\binits{K.~C.}}, \bauthor{\bsnm{Kim},~\bfnm{Katherine~J}\binits{K.~J.}}, \bauthor{\bsnm{Paulus},~\bfnm{Trena~M}\binits{T.~M.}}, \bauthor{\bsnm{Davies},~\bfnm{Evan~W}\binits{E.~W.}}, \bauthor{\bsnm{Stewart},~\bfnm{Katie~D}\binits{K.~D.}}, \bauthor{\bsnm{McComsey},~\bfnm{Grace~A}\binits{G.~A.}} \AND \bauthor{\bsnm{Fordyce},~\bfnm{Marshall~W}\binits{M.~W.}}
(\byear{2017}).
\btitle{Risks associated with antiretroviral treatment for human immunodeficiency virus (HIV): qualitative analysis of social media data and health state utility valuation}.
\bjournal{Quality of Life Research}
\bvolume{26}
\bpages{1785--1798}.
\end{barticle}
\endbibitem

\bibitem[\protect\citeauthoryear{Owen}{1980}]{DBOwen}
\begin{barticle}[author]
\bauthor{\bsnm{Owen},~\bfnm{D.~B.}\binits{D.~B.}}
(\byear{1980}).
\btitle{A table of normal integrals}.
\bjournal{Communications in Statistics - Simulation and Computation}
\bvolume{9}
\bpages{389-419}.
\bdoi{10.1080/03610918008812164}
\end{barticle}
\endbibitem

\bibitem[\protect\citeauthoryear{Pearl}{2001}]{Pearl_2001}
\begin{binproceedings}[author]
\bauthor{\bsnm{Pearl},~\bfnm{Judea}\binits{J.}}
(\byear{2001}).
\btitle{Direct and indirect effects}.
In \bbooktitle{Proceedings of the Seventeenth Conference on Uncertainty in Artificial Intelligence}.
\bseries{UAI'01}
\bpages{411–420}.
\bpublisher{Morgan Kaufmann Publishers Inc.}, \baddress{San Francisco, CA, USA}.
\end{binproceedings}
\endbibitem

\bibitem[\protect\citeauthoryear{Pearl}{2010}]{Pearl2010}
\begin{bmisc}[author]
\bauthor{\bsnm{Pearl},~\bfnm{Judea}\binits{J.}}
(\byear{2010}).
\btitle{An introduction to causal inference}.
\bdoi{10.2202/1557-4679.1203}
\end{bmisc}
\endbibitem

\bibitem[\protect\citeauthoryear{Pearl}{2011}]{Pearl2011}
\begin{bbook}[author]
\bauthor{\bsnm{Pearl},~\bfnm{Judea}\binits{J.}}
(\byear{2011}).
\btitle{Causality: Models, reasoning, and inference, second edition}.
\bdoi{10.1017/CBO9780511803161}
\end{bbook}
\endbibitem

\bibitem[\protect\citeauthoryear{Riddler et~al.}{2016}]{Riddler2016}
\begin{barticle}[author]
\bauthor{\bsnm{Riddler},~\bfnm{Sharon~A.}\binits{S.~A.}}, \bauthor{\bsnm{Aga},~\bfnm{Evgenia}\binits{E.}}, \bauthor{\bsnm{Bosch},~\bfnm{Ronald~J.}\binits{R.~J.}}, \bauthor{\bsnm{Bastow},~\bfnm{Barbara}\binits{B.}}, \bauthor{\bsnm{Bedison},~\bfnm{Margaret}\binits{M.}}, \bauthor{\bsnm{Vagratian},~\bfnm{David}\binits{D.}}, \bauthor{\bsnm{Vaida},~\bfnm{Florin}\binits{F.}}, \bauthor{\bsnm{Eron},~\bfnm{Joseph~J.}\binits{J.~J.}}, \bauthor{\bsnm{Gandhi},~\bfnm{Rajesh~T.}\binits{R.~T.}} \AND \bauthor{\bsnm{Mellors},~\bfnm{John~W.}\binits{J.~W.}}
(\byear{2016}).
\btitle{Continued slow decay of the residual plasma viremia level in HIV-1-infected adults receiving long-term antiretroviral therapy}.
\bjournal{Journal of Infectious Diseases}
\bvolume{213}.
\bdoi{10.1093/infdis/jiv433}
\end{barticle}
\endbibitem

\bibitem[\protect\citeauthoryear{Robins and Greenland}{1992}]{Robins1992}
\begin{barticle}[author]
\bauthor{\bsnm{Robins},~\bfnm{James~M.}\binits{J.~M.}} \AND \bauthor{\bsnm{Greenland},~\bfnm{Sander}\binits{S.}}
(\byear{1992}).
\btitle{Identifiability and exchangeability for direct and indirect effects}.
\bjournal{Epidemiology}
\bvolume{3}.
\bdoi{10.1097/00001648-199203000-00013}
\end{barticle}
\endbibitem

\bibitem[\protect\citeauthoryear{Schinazi}{2022}]{Schinazi2022}
\begin{binbook}[author]
\bauthor{\bsnm{Schinazi},~\bfnm{Rinaldo~B.}\binits{R.~B.}}
(\byear{2022}).
\btitle{The Bivariate Normal Distribution}
In \bbooktitle{Probability with Statistical Applications}
\bpages{209--217}.
\bpublisher{Springer International Publishing}, \baddress{Cham}.
\bdoi{10.1007/978-3-030-93635-8_19}
\end{binbook}
\endbibitem

\bibitem[\protect\citeauthoryear{Scully et~al.}{2022}]{Scully2022}
\begin{barticle}[author]
\bauthor{\bsnm{Scully},~\bfnm{Eileen~P.}\binits{E.~P.}}, \bauthor{\bsnm{Aga},~\bfnm{Evgenia}\binits{E.}}, \bauthor{\bsnm{Tsibris},~\bfnm{Athe}\binits{A.}}, \bauthor{\bsnm{Archin},~\bfnm{Nancie}\binits{N.}}, \bauthor{\bsnm{Starr},~\bfnm{Kate}\binits{K.}}, \bauthor{\bsnm{Ma},~\bfnm{Qing}\binits{Q.}}, \bauthor{\bsnm{Morse},~\bfnm{Gene~D.}\binits{G.~D.}}, \bauthor{\bsnm{Squires},~\bfnm{Kathleen~E.}\binits{K.~E.}}, \bauthor{\bsnm{Howell},~\bfnm{Bonnie~J.}\binits{B.~J.}}, \bauthor{\bsnm{Wu},~\bfnm{Guoxin}\binits{G.}}, \bauthor{\bsnm{Hosey},~\bfnm{Lara}\binits{L.}}, \bauthor{\bsnm{Sieg},~\bfnm{Scott~F.}\binits{S.~F.}}, \bauthor{\bsnm{Ehui},~\bfnm{Lynsay}\binits{L.}}, \bauthor{\bsnm{Giguel},~\bfnm{Francoise}\binits{F.}}, \bauthor{\bsnm{Coxen},~\bfnm{Kendyll}\binits{K.}}, \bauthor{\bsnm{Dobrowolski},~\bfnm{Curtis}\binits{C.}}, \bauthor{\bsnm{Gandhi},~\bfnm{Monica}\binits{M.}}, \bauthor{\bsnm{Deeks},~\bfnm{Steve}\binits{S.}}, \bauthor{\bsnm{Chomont},~\bfnm{Nicolas}\binits{N.}},
  \bauthor{\bsnm{Connick},~\bfnm{Elizabeth}\binits{E.}}, \bauthor{\bsnm{Godfrey},~\bfnm{Catherine}\binits{C.}}, \bauthor{\bsnm{Karn},~\bfnm{Jonathan}\binits{J.}}, \bauthor{\bsnm{Kuritzkes},~\bfnm{Daniel~R.}\binits{D.~R.}}, \bauthor{\bsnm{Bosch},~\bfnm{Ronald~J.}\binits{R.~J.}} \AND \bauthor{\bsnm{Gandhi},~\bfnm{Rajesh~T.}\binits{R.~T.}}
(\byear{2022}).
\btitle{Impact of Tamoxifen on Vorinostat-Induced Human Immunodeficiency Virus Expression in Women on Antiretroviral Therapy: AIDS Clinical Trials Group A5366, The MOXIE Trial}.
\bjournal{Clinical infectious diseases : an official publication of the Infectious Diseases Society of America}
\bvolume{75}.
\bdoi{10.1093/cid/ciac136}
\end{barticle}
\endbibitem

\bibitem[\protect\citeauthoryear{UNAIDS}{2024}]{unaids2024Global}
\begin{bmisc}[author]
\bauthor{\bsnm{UNAIDS}}
(\byear{2024}).
\btitle{2024 global {A}{I}{D}{S} report — {T}he {U}rgency of {N}ow: {A}{I}{D}{S} at a {C}rossroads --- unaids.org}.
\bhowpublished{\url{https://www.unaids.org/en/resources/documents/2024/global-aids-update-2024}}.
\bnote{[Accessed 01-09-2024]}.
\end{bmisc}
\endbibitem

\bibitem[\protect\citeauthoryear{Valeri, Lin and Vanderweele}{2014}]{Valeri2014_ME}
\begin{barticle}[author]
\bauthor{\bsnm{Valeri},~\bfnm{Linda}\binits{L.}}, \bauthor{\bsnm{Lin},~\bfnm{Xihong}\binits{X.}} \AND \bauthor{\bsnm{Vanderweele},~\bfnm{Tyler~J.}\binits{T.~J.}}
(\byear{2014}).
\btitle{Mediation analysis when a continuous mediator is measured with error and the outcome follows a generalized linear model}.
\bjournal{Statistics in Medicine}
\bvolume{33}
\bpages{4875-4890}.
\bdoi{10.1002/sim.6295}
\end{barticle}
\endbibitem

\bibitem[\protect\citeauthoryear{Vanderweele}{2013}]{Vanderweele2013}
\begin{barticle}[author]
\bauthor{\bsnm{Vanderweele},~\bfnm{Tyler~J.}\binits{T.~J.}}
(\byear{2013}).
\btitle{A three-way decomposition of a total effect into direct, indirect, and interactive effects}.
\bjournal{Epidemiology}
\bvolume{24}.
\bdoi{10.1097/EDE.0b013e318281a64e}
\end{barticle}
\endbibitem

\bibitem[\protect\citeauthoryear{Vanderweele}{2015}]{Vanderweele2015}
\begin{bbook}[author]
\bauthor{\bsnm{Vanderweele},~\bfnm{T.~J.}\binits{T.~J.}}
(\byear{2015}).
\btitle{Explanation in causal inference: Methods for mediation and interaction}.
\bpublisher{Oxford University Press}.
\end{bbook}
\endbibitem

\end{thebibliography}
